\documentclass[lettersize,journal]{IEEEtran}
\usepackage{amsmath,amsfonts}
\usepackage{algorithm}
\usepackage{algpseudocode}
\usepackage{array}
\usepackage[caption=false,font=normalsize,labelfont=sf,textfont=sf]{subfig}
\usepackage{textcomp}
\usepackage{stfloats}
\usepackage{url}
\usepackage{verbatim}
\usepackage{graphicx}
\usepackage{tabularx}
\usepackage{multirow}
\usepackage{cite}
\usepackage{array}
\usepackage{makecell}
\usepackage[colorlinks, linkcolor=blue, anchorcolor=blue, citecolor=blue, urlcolor=blue]{hyperref}
\usepackage{orcidlink}
\def\BibTeX{{\rm B\kern-.05em{\sc i\kern-.025em b}\kern-.08em
    T\kern-.1667em\lower.7ex\hbox{E}\kern-.125emX}}
\usepackage{balance}
\begin{document}
\title{Measurement-Based Characterization and Statistical Modeling of 6G Urban Low-Altitude A2G Channels across FR1 and FR3}

\author{Bin Ao\orcidlink{0009-0006-6509-6455},~\IEEEmembership{Graduate Student Member,~IEEE,}
        Jianhua Zhang\orcidlink{0000-0002-6492-3846},~\IEEEmembership{Fellow,~IEEE,}
        
        Pan Tang\orcidlink{0000-0003-0432-7361},~\IEEEmembership{Member,~IEEE,}        
        Peijie Liu\orcidlink{0009-0005-2196-7215},~\IEEEmembership{Graduate Student Member,~IEEE,}
        
        Boyang He\orcidlink{0009-0005-4717-8556},~\IEEEmembership{Graduate Student Member,~IEEE,}
        and~Hao Zheng\orcidlink{0009-0000-7502-4662},~\IEEEmembership{Graduate Student Member,~IEEE,}
\thanks{
	This work was supported in part by the National Natural Science Foundation of China (NCFC) under Grant 62571053, Grant 62525101 and Grant 62341128;
	in part by the National Key Research and Development Program of China under grant 2023YFB2904805, 
	and in part by the Beijing Municipal Natural Fund under grant L243002.
	\textit{(Corresponding authors: Jianhua Zhang.)}}         
\thanks{Bin Ao, Jianhua Zhang, Pan Tang, Peijie Liu, Boyang He, and Hao Zheng are with the State Key Laboratory of Networking and Switching Technology, Beijing University of Posts and Telecommunications, Beijing 100876, China (e-mail: binao@bupt.edu.cn; jhzhang@bupt.edu.cn; tangpan27@bupt.edu.cn; liupj@bupt.edu.cn; heboyang@bupt.edu.cn; zheng.hao@bupt.edu.cn).}}



\maketitle

\begin{abstract}
Unmanned aerial vehicle (UAV) communications have been recognized as a key component of future sixth-generation (6G) space-air-ground-sea integrated networks. Accurate characterization and modeling of air-to-ground (A2G) channels are essential for the design and optimization of low-altitude communication systems. This paper presents a wideband A2G channel measurement campaign in an urban environment at 2.85 and 4.6~GHz in FR1 and 7.25~GHz in the FR3 frequency band, each with a bandwidth of 250~MHz. To enable reliable line-of-sight (LoS) and non-line-of-sight (NLoS) propagation state identification, a weakly supervised method is developed by fusing geometric priors, channel features, and spatial consistency constraints. Furthermore, based on the measured data, A2G channel characteristics are extracted and analyzed under LoS/NLoS conditions across different frequency bands, including path loss (PL), shadow fading (SF), power delay profile, root-mean-square delay spread (RMS-DS), and Rician $K$-factor. The results show that the close-in model fits the measured PL more accurately than the 3GPP reference model, and that NLoS propagation leads to larger path loss exponents and stronger SF than LoS propagation. For channel delay characteristics, higher-frequency channels exhibit fewer effective MPCs and weaker delay dispersion, indicating increased channel sparsity. Specifically, the mean RMS-DS under LoS conditions decreases from 93.11 to 46.84~ns, while the mean Rician $K$-factor increases from 9.16 to 12.88~dB. The statistical results further show that the RMS-DS and the Rician $K$-factor can be well characterized by lognormal and normal distributions, respectively. Moreover, the movement of the receiver in a complex scattering environment intensifies the spatial non-stationarity of the A2G channel.

\end{abstract}


\section{Introduction}

\IEEEPARstart{W}{ith} the rapid development of wireless communications, sixth-generation (6G) aims to construct a space-air-ground-sea integrated network (SAGSIN) to achieve ubiquitous connectivity and seamless coverage \cite{bacv1,bacv2,bacv3}. As a critical link between the aerial and terrestrial networks, low-altitude communications exhibit increasing strategic significance and economic value \cite{lowalteconomecy1}. Owing to their high mobility, flexible deployment, and low cost, unmanned aerial vehicles (UAVs) have emerged as key platforms for low-altitude communications and offer broad potential in representative scenarios, including emergency communication support, temporary hotspot coverage enhancement, and communication assistance for vehicular-to-everything (V2X) networks \cite{appscenario}. However, as the scale of low-altitude communication services continues to expand, the existing spectrum resources are insufficient to support the diverse service requirements of future systems. Currently, the protected dedicated spectrum allocated for UAV communications remains limited. The International Telecommunication Union (ITU) primarily designates the 960–1164 MHz and 5030–5091 MHz bands for Control and Non-Payload Communications (CNPC) \cite{ITU}. Meanwhile, research and standardization efforts regarding potential candidate bands are continuously advancing for future wireless communication systems \cite{midbandv1,midbandv2}. According to the ITU-R agenda items for WRC-27, the 4.4--4.8~GHz band in FR1 and the 7.125--8.4~GHz and 14.8--15.35~GHz bands in the FR3 candidate range have been identified as candidate IMT bands for further investigation \cite{ITUWRC23,midbandv3,midbandv4}. In this context, conducting multi-frequency air-to-ground (A2G) channel measurements, characterization, and modeling is essential to support the design and optimization of low-altitude communication systems.

When conducting A2G channel measurements, UAVs serve as the carrier platforms and are primarily categorized into fixed-wing and rotary-wing configurations. These configurations exhibit significant differences in structural design and flight characteristics, directly impacting the applicable scope of channel modeling. Specifically, fixed-wing UAVs typically feature higher flight altitudes, longer cruising ranges, and greater payload capacities. In contrast, rotary-wing UAVs possess excellent hovering capabilities and flexibility, providing distinct advantages for short-distance and high-precision measurements in low-altitude and complex physical environments such as urban canyons.

To date, extensive measurement and modeling studies have been conducted on A2G channels\cite{modelv1,modelv2,modelv3}. Most existing works focus on low-frequency bands, whereas field measurements at higher frequencies, such as the millimeter-wave (mmWave), remain relatively limited. In the sub-6 GHz band, \cite{A2G968M} performs channel measurements in suburban and near-suburban scenarios at 968 MHz and 5.06 GHz by using a fixed-wing UAV, analyzes the path loss (PL), root-mean-square delay spread (RMS-DS), and Rician $K$ factor, and establishes a tapped delay line (TDL) model. \cite{A2G2p85v1,A2G2p85v2} evaluate the differences in large-scale and small-scale fading characteristics at different flight altitudes in suburban scenarios under long-time-evolution (LTE) networks. The results show that the channel characteristics are still affected by the ground scattering environment even when the UAV operates at a relatively high altitude. Similarly, \cite{A2GPLheight} observed an increasing trend in the path loss exponent (PLE) with UAV flight altitude and further investigated the effect of vegetation blockage on PL. \cite{A2GSIMO} deployed a 128-antenna array at the ground terminal to characterize the time-space distribution of MPCs in A2G channels. The study shows that tree crown blockage has a significant impact on A2G communications and proposes a PL model for forest scenarios. To investigate the spatial distribution of multipath components (MPCs) in A2G channels, \cite{A2G3p5cam} develops a single-input multiple-output (SIMO) channel sounder and conducts measurements at 3.5 GHz in a campus scenario. In addition, \cite{A2G70M} designs an expert-assisted attention network that incorporates physical channel prior knowledge and enables real-time and high-accuracy prediction of A2G channel PL in urban scenarios.

\newcolumntype{Y}{>{\centering\arraybackslash}X} 
\renewcommand{\tabularxcolumn}[1]{m{#1}} 
\begin{table*}[t]
	\renewcommand{\arraystretch}{1.3} 
	\caption{Related works on A2G channel measurement and modeling.}
	\begin{center}
		\begin{tabularx}{\linewidth}{>{\centering\arraybackslash}m{1.2cm}|>{\centering\arraybackslash}m{1.5cm}|>{\centering\arraybackslash}m{2.2cm}|>{\centering\arraybackslash}m{1.4cm}|>{\centering\arraybackslash}m{1.6cm}|Y}
			\hline \hline
			\textbf{Ref.} & \textbf{Scenario} & \textbf{Center Frequency} & \textbf{Bandwidth} & \textbf{UAV Type} & \textbf{Channel Characteristics} \\
			\hline
			\cite{A2G968M} & Suburban, Near-urban & \makecell{968 MHz, \\ 5060 MHz} & \makecell{5 MHz, \\ 50 MHz} & Fixed-wing & PL, RMS-DS, Rician $K$ factor, TDL model \\
			\hline
			\cite{A2G2p85v1,A2G2p85v2} & Suburban & 2.585 GHz & 18 MHz & Rotor-wing & PDP, Doppler frequency spectrum, PL, SF, Rician $K$ factor, RMS-DS, RMS Doppler frequency spread \\
			\hline
			\cite{A2GPLheight} & Suburban & 2.61 GHz & 18 MHz & Fixed-wing & PL \\
			\hline
			\cite{A2GSIMO} & Urban & 3.5 GHz & 46 MHz & Rotor-wing & PDP, PAS \\
			\hline
			\cite{A2G3p5cam} & Campus & 3.5 GHz & 100 MHz & Rotor-wing & PDP, PL, SF, Rician $K$ factor, AoA, EoA \\
			\hline
			\cite{A2G70M} & Urban & 70 MHz & 5 MHz & Rotor-wing & PL, SF, RMS-DS, Doppler shift, Rician $K$ factor, TDL model, Neural-enhanced PL prediction model \\
			\hline
			\cite{A2G6p5v1} & Lake, Rural, Hilly, Suburban & 6.5 GHz & 500 MHz & Rotor-wing & PDP, Rician $K$ factor, RMS-DS \\
			\hline
			\cite{smartfactory} & Smart factory & 4.2 GHz & 2.2 GHz & Rotor-wing & PL, APDP, RMS-DS \\
			\hline
			\cite{A2G27a38G} & Urban & 27, 38 GHz & 1 MHz & Rotor-wing & PAP \\
			\hline
			\cite{A2G60G} & Suburban & 60 GHz &  & Rotor-wing & Path loss model, Sub-optimal beam selection \\
			\hline
			\cite{A2G28Gv1} & Airport & 28 GHz &  & Rotor-wing & PL \\
			\hline
			\cite{A2G26Gv2} & Rural & 26 GHz & 1 GHz & Rotor-wing & PL, SF, PADP, RMS-DS, Rician $K$ factor, ASA, ESA, Channel sparsity, Correlation coefficient, Decorrelation distance, Cluster parameters \\
			\hline
			\cite{A2G28Gv2} & Rural & 28 GHz & 420 MHz & Fixed-wing & PDP, Doppler frequency spectrum, PL, RMS-DS, Rician $K$ factor, RMS Doppler spread \\
			\hline \hline
		\end{tabularx}
		\label{tab:related work}
	\end{center}
\end{table*}

For wideband and higher-frequency A2G channels, \cite{A2G6p5v1} comparatively evaluate the Rician $K$ factor and RMS-DS of ultra-wideband channels at 6.5 GHz in different scenarios. For smart factory applications of UAVs, \cite{smartfactory} conducted ultra-wideband A2G channel measurements over the 3.1--5.3 GHz frequency band. The results show that the ground scattering environment has a significant effect on the channel characteristics. In the mmWave band, \cite{A2G27a38G} obtains the power-azimuth profile (PAP) at 27 and 38 GHz under different elevation angles in an urban street canyon scenario by rotating a horn antenna mounted on the UAV through servo control. \cite{A2G60G} investigated the effects of UAV body blockage and attitude variations on suboptimal beam selection, and developed a stochastic channel model for PL evaluation. To mitigate UAV body effects, \cite{A2G28Gv1} proposed an antenna radiation pattern calibration method and further established a PL model. \cite{A2G26Gv2} conduct measurements at 26 GHz in rural railway scenarios by using omnidirectional antennas and phased-array antennas, respectively, and provide a comprehensive characterization of the large-scale parameters (LSPs). The results show that mmWave A2G channels exhibit significant sparsity. In addition, \cite{A2G28Gv2} conducts channel measurements at 28 GHz in a rural scenario and establishes a stochastic channel model by incorporating the flight altitude and attitude of the UAV. Table~\ref{tab:related work} summarizes the A2G channel measurement campaigns and related research.

In general, numerous A2G channel measurements primarily focus on single-band scenarios with limited bandwidth. In contrast, studies on multi-frequency channel characteristics are more common in terrestrial communication networks \cite{bSub-6,bSparsity,bTHz}. Moreover, restricted by flight regulations across different countries and regions, A2G channel measurements in urban scenarios remain relatively limited. For candidate communication bands spanning FR1 and FR3, employing a unified channel measurement system, consistent scenario planning, and identical data post-processing methods provides a fair baseline for characterizing and modeling multi-frequency channel properties. To fill these gaps, this paper presents multi-frequency A2G channel measurements at 2.85 GHz, 4.6 GHz, and 7.25 GHz in an urban scenario to comprehensively compare and analyze the LSPs across different frequency bands. The main contributions are summarized as follows:
\begin{itemize}
    \item A multi-frequency wideband A2G channel measurement campaign is conducted at 2.85, 4.6, and 7.25~GHz in an urban scenario, covering both line-of-sight (LoS) and non-line-of-sight (NLoS) conditions. By deploying a UAV-mounted transmitter hovering at an altitude of approximately 150~m and a vehicle-mounted receiver moving along a consistent urban route, a large amount of continuous channel impulse response (CIR) data is acquired.    
    \item A weakly supervised LoS/NLoS propagation state identification method is proposed for A2G channel measurements. By integrating geometry-based Fresnel clearance priors, channel features, and hidden Markov model (HMM)-based smoothing, the proposed method obtains propagation state sequences that balance identification accuracy and spatial consistency.
    \item Based on the measured data and identified propagation states, the large-scale and small-scale fading characteristics of multi-frequency A2G channels are comprehensively analyzed and statistically modeled. The investigated characteristics include PL, SF, RMS-DS, Rician $K$-factor, cross-correlations among large-scale parameters, and spatial non-stationarity of the channel.   
\end{itemize}

The remainder of this paper is organized as follows. Section II introduces the multi-frequency A2G channel measurement campaign. Section III presents a weakly supervised method for LoS/NLoS propagation state identification. Section IV analyzes and statistically models the large-scale and small-scale channel characteristics, as well as the cross-correlation and spatial non-stationarity of multi-frequency A2G channels. Finally, Section V concludes the paper.

\section{Multi-Frequency A2G Channel Measurement}

This section presents the multi-frequency A2G channel measurement campaign in the urban scenario, including the measurement system architecture, scenario planning, and data post-processing method. The following subsections describe these components in detail.

\subsection{Measurement System}

\begin{figure}[!t]
	\centering
	\subfloat[]{\includegraphics[width=1\columnwidth]{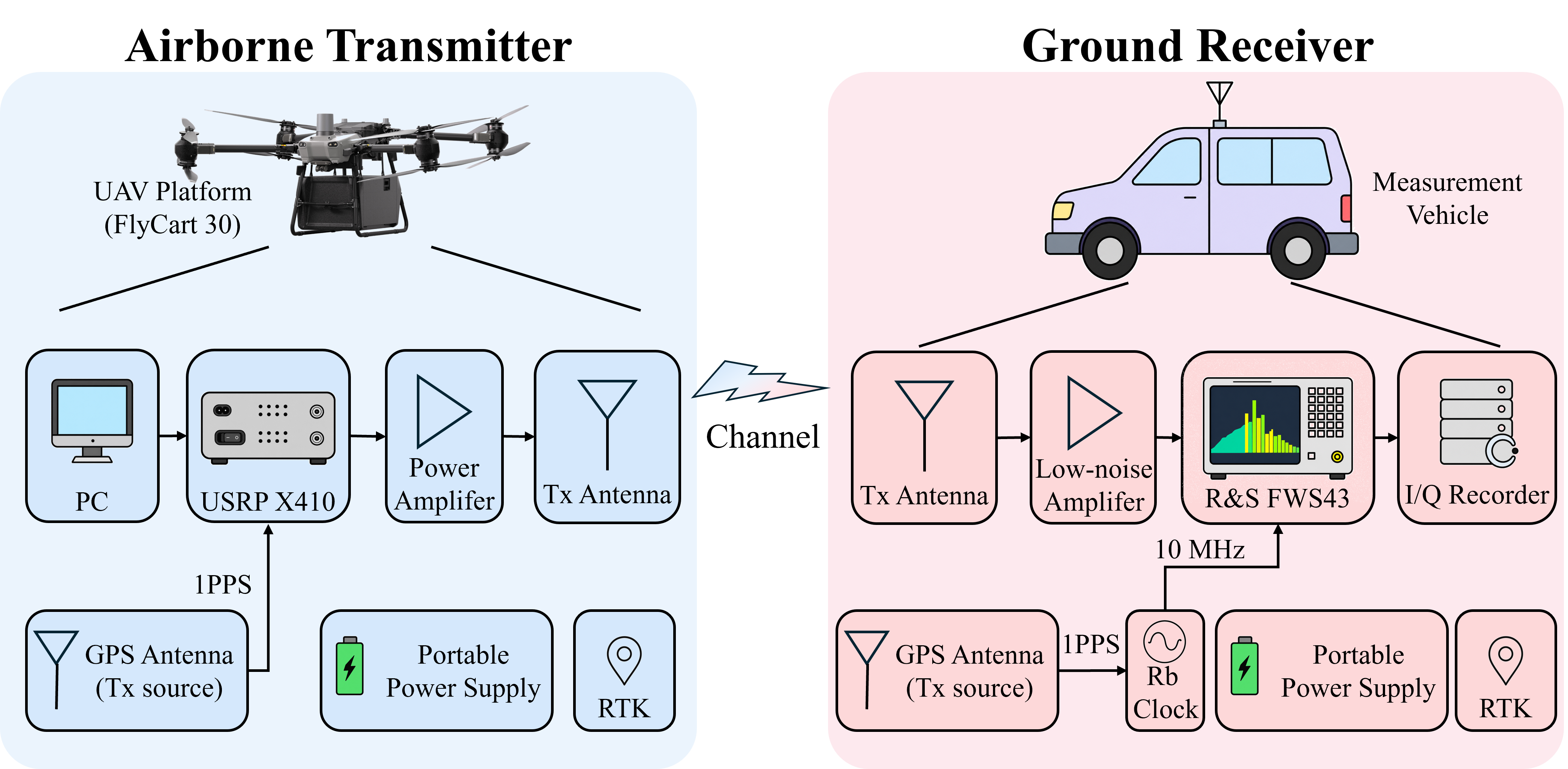}%
		\label{fig:system_arch}}
	\hfil
	\subfloat[]{\includegraphics[width=0.55\columnwidth]{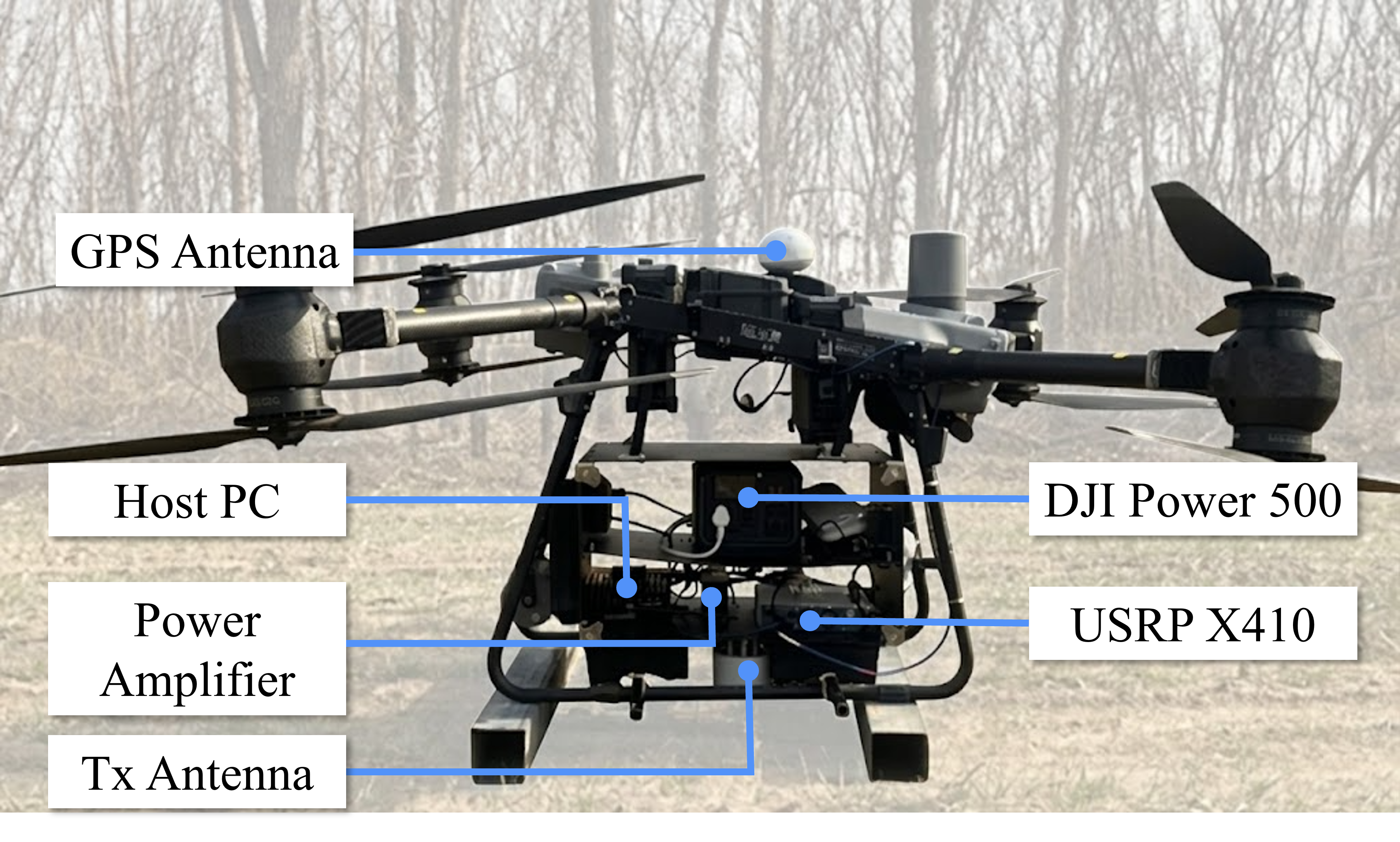}%
		\label{fig:equipment_tx}}
	\hfil
	\subfloat[]{\includegraphics[width=0.45\columnwidth]{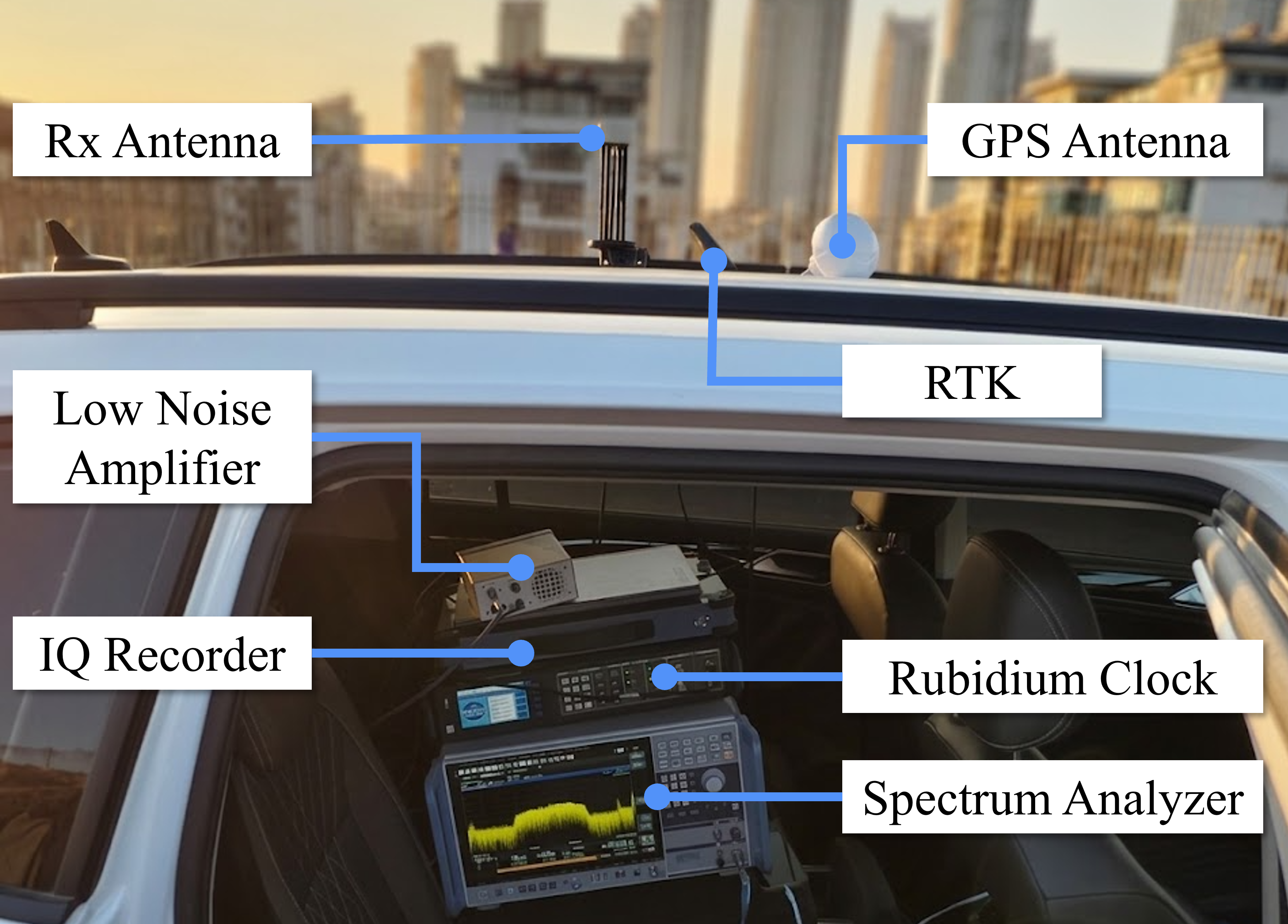}%
		\label{fig:equipment_rx}}
	\caption{Overview of the A2G channel measurement system: (a) Block diagram of the measurement system, (b) Key equipment of the UAV-mounted Tx platform, and (c) Key equipment of the vehicle-mounted Rx platform.}
	\label{fig:equipment}
\end{figure}

Fig.~\ref{fig:equipment}(a) illustrates the architecture of the developed wideband A2G channel measurement system. The measurement platform consists of a UAV-mounted airborne transmitter (Tx) and a vehicle-based ground receiver (Rx). The system operates on the time-domain spread spectrum sliding correlation principle. By transmitting pseudo-noise (PN) sequences with favorable autocorrelation properties, it achieves a high dynamic range for channel sounding, thereby enabling accurate acquisition of time-varying CIRs.

The DJI FlyCart 30 UAV was employed as the airborne platform, as shown in Fig.~\ref{fig:equipment}(b). The Tx payload mainly consists of a USRP X410, a power amplifier (PA), a host computer, a GPS antenna for synchronization, an RTK-GNSS positioning module, an omnidirectional antenna, and a portable power supply. All onboard components are interconnected via RF cables and rigidly secured inside a customized cargo box. Under the control of the host computer, the USRP X410 continuously transmits a 511-chip PN sequence with BPSK modulation at the target frequency bands. The sounding signal is amplified by a PA and radiated into the surrounding environment through an omni-antenna. To reduce structural blockage and electromagnetic interference from the UAV body and rotors, the Tx antenna is mounted beneath the UAV. The RTK-GNSS module is mounted on the top of the UAV to record the spatial coordinates of the airborne platform.

The ground-based Rx is deployed inside a measurement vehicle, as shown in Fig.~\ref{fig:equipment}(c). Its main components include a Rohde~\&~Schwarz (R\&S) FSW43 spectrum analyzer, a low-noise amplifier (LNA), an in-phase/quadrature (I/Q) data recorder, a rubidium (Rb) clock, a GPS antenna for synchronization, an RTK-GNSS positioning module, an omni-antenna, and a portable power supply. The received RF signal is first amplified by the LNA and then monitored in real time using the spectrum analyzer. Under the control of the host computer, the received I/Q data streams are recorded at high speed by the I/Q recorder.

\begin{table}[t]
	\renewcommand{\arraystretch}{1.3}
	\caption{The A2G Channel Measurement Configurations.}
	\begin{center}
		\begin{tabularx}{\linewidth}{>{\centering\arraybackslash}X|>{\centering\arraybackslash}X}
			\hline \hline
			\textbf{Configuration} & \textbf{Value} \\
			\hline
			Center frequency, $f_c$ & 2.85 / 4.6 / 7.25~GHz \\
			\hline
			Bandwidth, $B$ & 250~MHz \\
			\hline
			Sounding signal & PN9 sequence \\
			\hline
			Tx antenna gain & -1.72 / 1.6 / 2.52~dBi \\
			\hline
			Rx antenna gain & 1 / 0.6 / 2.3~dBi \\
			\hline
			UAV flight altitude & $\sim$150~m \\
			\hline
			Rx antenna height & $\sim$1.8~m \\
			\hline
			Longitude & 120$^\circ$\,17$'$\,07.14$''$\,E \\
			\hline
			Latitude & 35$^\circ$\,59$'$\,04.23$''$\,N \\
			\hline \hline
		\end{tabularx}
		\label{tab:system_param}
	\end{center}
\end{table}

Since the Tx and Rx are spatially separated and independently mobile in the A2G scenario, cable-based synchronization is impractical. Therefore, a GPS-disciplined Rb clock synchronization scheme is employed. Specifically, the Rb clock provides a stable 10~MHz reference for the RF components, while the GPS antenna provides a synchronized pulse-per-second (PPS) trigger for system-level synchronization between the Tx and Rx. This synchronization architecture ensures the timing consistency required for reliable A2G channel measurements in dynamic scenarios. The key configurations of the overall measurement system are summarized in Table~\ref{tab:system_param}.

\subsection{Measurement Scenario}

\begin{figure}[!t]
	\centering
	\subfloat[]{\includegraphics[width=0.67\columnwidth]{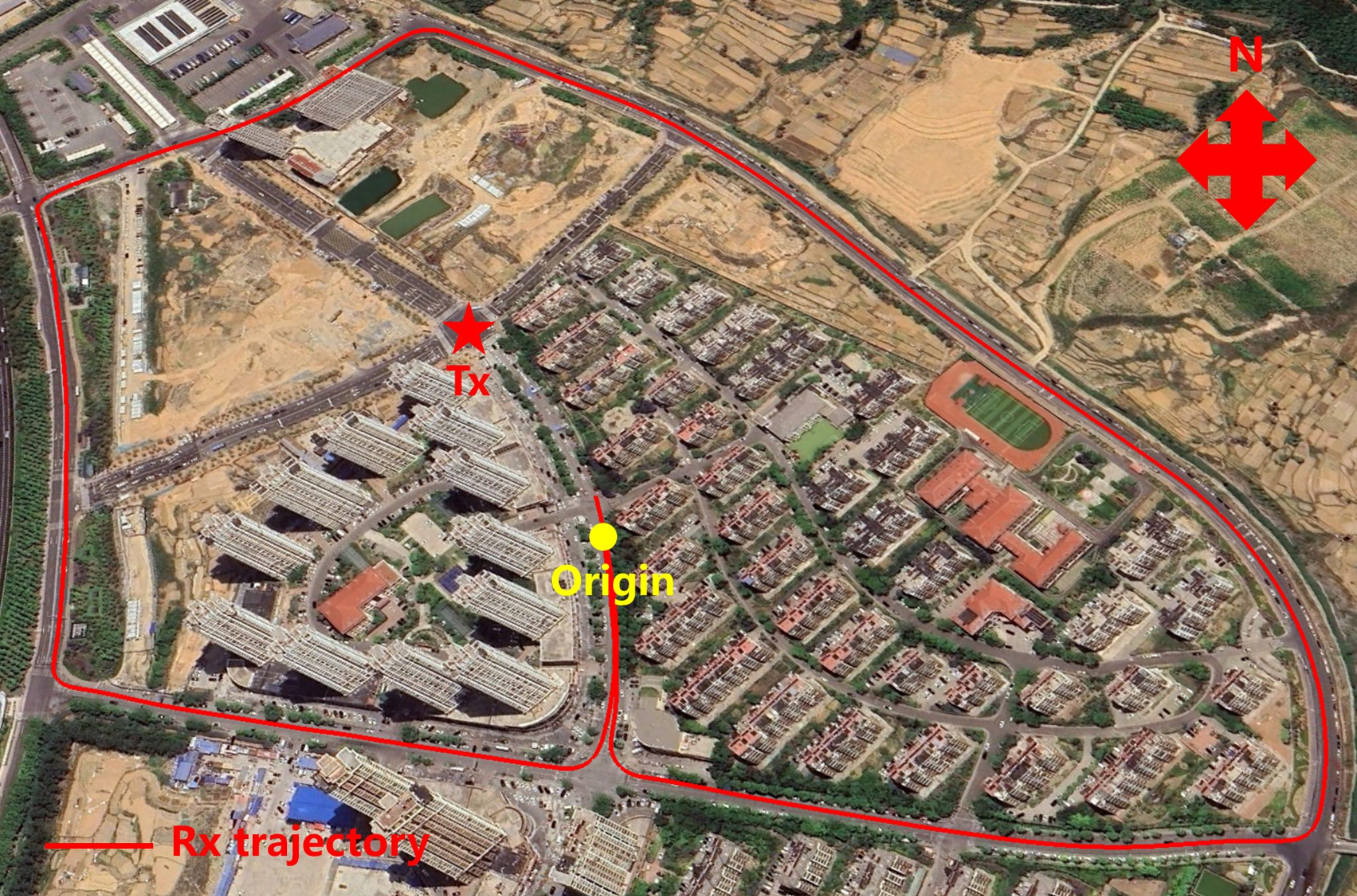}%
		\label{fig:scenario_sat}}
	\hfil
	\subfloat[]{\includegraphics[width=0.33\columnwidth]{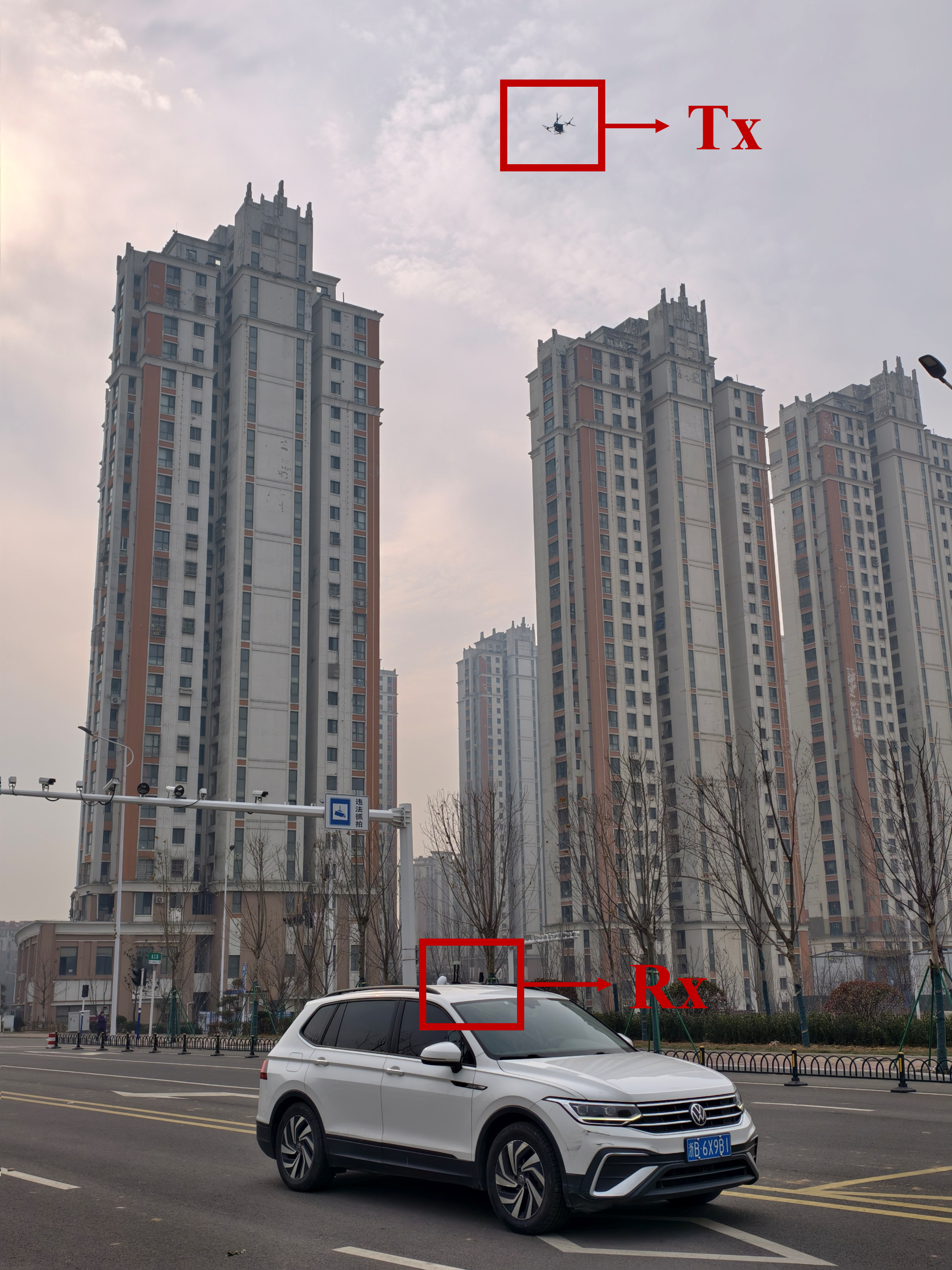}%
		\label{fig:scenario_photo}}
	\caption{A2G channel measurement scenario: (a) satellite view of the measurement scenario adapted from Google Maps, showing the UAV hovering position (red star), Rx starting point (yellow dot), and Rx trajectory (red line); (b) ground-level photograph of the measurement scenario.}
	\label{fig:scenario}
\end{figure}

The A2G channel measurements were conducted in Qingdao, Shandong Province, China (35.9845$^\circ$~N, 120.2853$^\circ$~E). The satellite view and ground-level photograph of the measurement scenario are shown in Fig.~\ref{fig:scenario}(a) and (b), respectively. The selected area represents a typical urban propagation environment. In the southern part of the scenario, both high-rise and low-rise buildings are densely distributed, with average heights of approximately 100~m and 30~m, respectively. In contrast, the northern part is relatively open with fewer buildings. In addition, roadside objects such as lampposts, trees, and vehicles are distributed along the streets and contribute additional scattering paths to the A2G channel.

Measurements were conducted at 2.85, 4.6, and 7.25~GHz, each with a bandwidth of 250~MHz. The 4.6~GHz and 7.25~GHz bands were selected from candidate frequency ranges under consideration in ITU WRC-27, while 2.85~GHz was included as a lower-frequency comparison band. During the measurement campaign, the UAV hovered at the red-star location shown in Fig.~\ref{fig:scenario}(a), with an altitude of approximately 150~m above ground level, far exceeding the height of all surrounding scatterers. The Rx antenna was mounted on the roof of a vehicle at a height of approximately 1.8~m. For all three frequency bands, the vehicle moved along the predefined urban route indicated by the red trajectory in Fig.~\ref{fig:scenario}(a), with a speed of approximately 10--40~km/h. During the vehicle movement, blockage of the direct path by surrounding buildings led to varying LoS/NLoS propagation conditions along the route.

\subsection{Data Post Processing}

The measured raw data contain not only the channel response but also the system response introduced by the transceiver hardware and RF cables. To remove this system response, a back-to-back (B2B) calibration is conducted prior to the field measurements. Specifically, the Tx and Rx are directly connected through a known attenuator to obtain the calibration signal. The CIR of the \(i\)-th snapshot is then obtained as

\begin{equation}
	h_i(\tau)
	=
	\mathcal{F}^{-1}\!\left\{
	\frac{
		\mathcal{F}\!\left\{y_i(\tau)\right\}
	}{
		\mathcal{F}\!\left\{y_{\mathrm{cal}}(\tau)\right\}
	}
	\right\},
	\label{eq:calibrated_cir}
\end{equation}
where \(y_i(\tau)\) denotes the received signal extracted from the raw I/Q data, \(y_{\mathrm{cal}}(\tau)\) denotes the B2B calibration response including the attenuator, and \(\mathcal{F}\{\cdot\}\) represents the Fourier transform.

The PDP is conventionally used to characterize the power distribution of MPCs over propagation delays. The PDP of the \(i\)-th snapshot is calculated from the calibrated CIR as

\begin{equation}
	P_i(\tau)=\left|h_i(\tau)\right|^2,
	\label{eq}
\end{equation}
Based on the measured PDP, a peak detection algorithm is employed for MPC extraction. Specifically, delay bins whose powers exceed the average noise floor by 6~dB are identified as valid multipath components \cite{b6dB}.

\begin{figure}[!t]
	\centering
	\includegraphics[width=1\columnwidth]{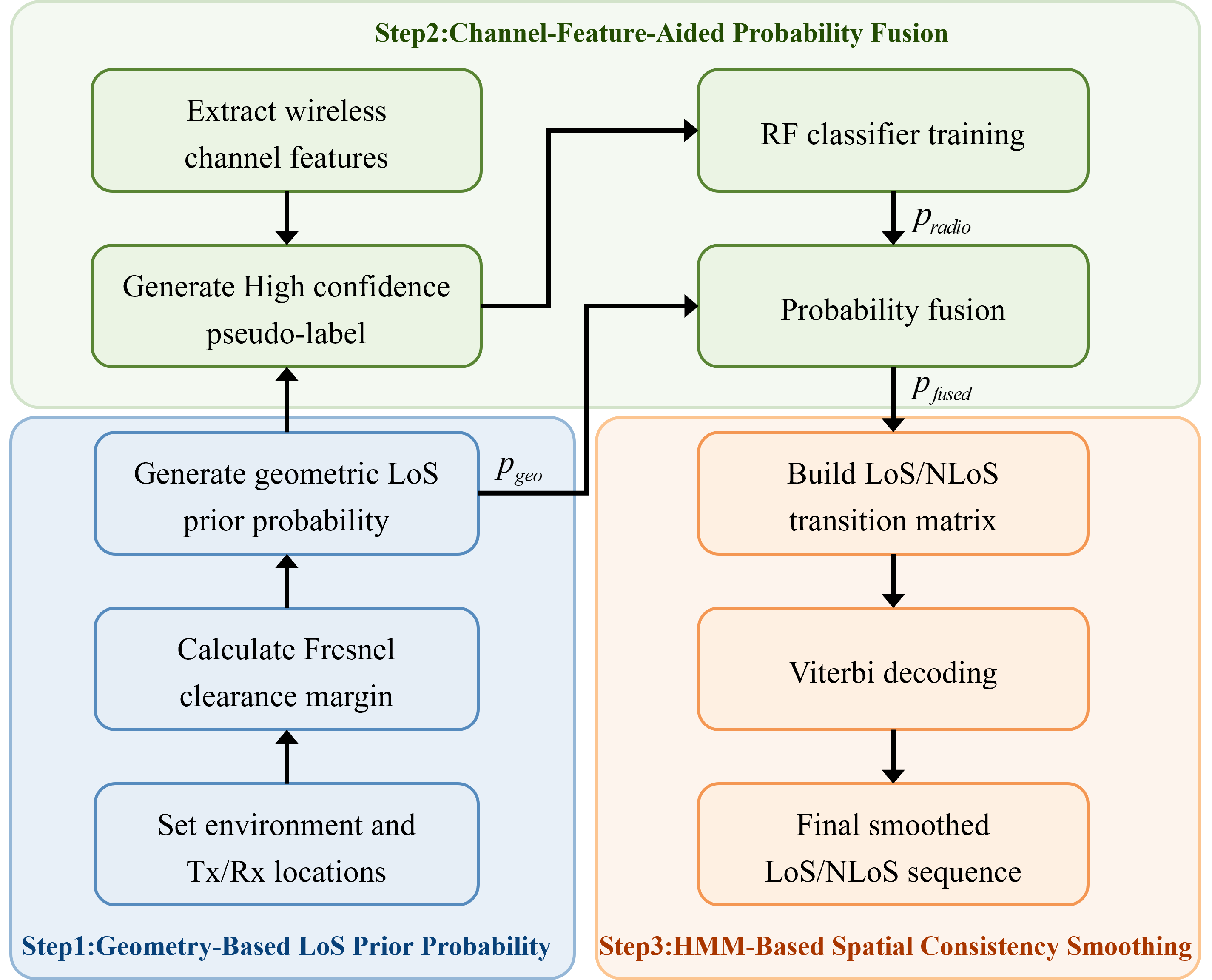}
	\caption{Flowchart of the proposed weakly supervised LoS/NLoS identification method.}
	\label{fig:LoS/NLoS identification flowchart}
\end{figure}

\section{LoS/NLoS Propagation State Identification for A2G Channels}

Accurate LoS/NLoS state identification is essential for A2G channel characterization and can also provide important support for UAV positioning \cite{UAVpositionv1,UAVpositionv2}. In outdoor channel measurements, obtaining snapshot level reference labels often requires manual annotation based on known Tx/Rx geometries or synchronized video recordings, which is difficult to extend to large scale channel measurement campaigns. To address this issue, this paper proposes a weakly supervised LoS/NLoS state identification framework that combines imperfect geometric priors and channel features, as shown in Fig.~\ref{fig:LoS/NLoS identification flowchart}. Specifically, a geometric prior based on Fresnel clearance is first constructed to provide a LoS probability rather than a hard LoS/NLoS decision. Then, high confidence LoS and NLoS samples are selected as pseudo labels to train a classifier using channel features, whose output is fused with the geometric prior to refine the LoS/NLoS decision. Finally, a HMM-based smoothing strategy is introduced to achieve spatially consistent propagation state identification.

\subsection{Geometry-Based LoS Prior Probability}

\begin{figure}[!t]
	\centering
	\includegraphics[width=0.8\columnwidth]{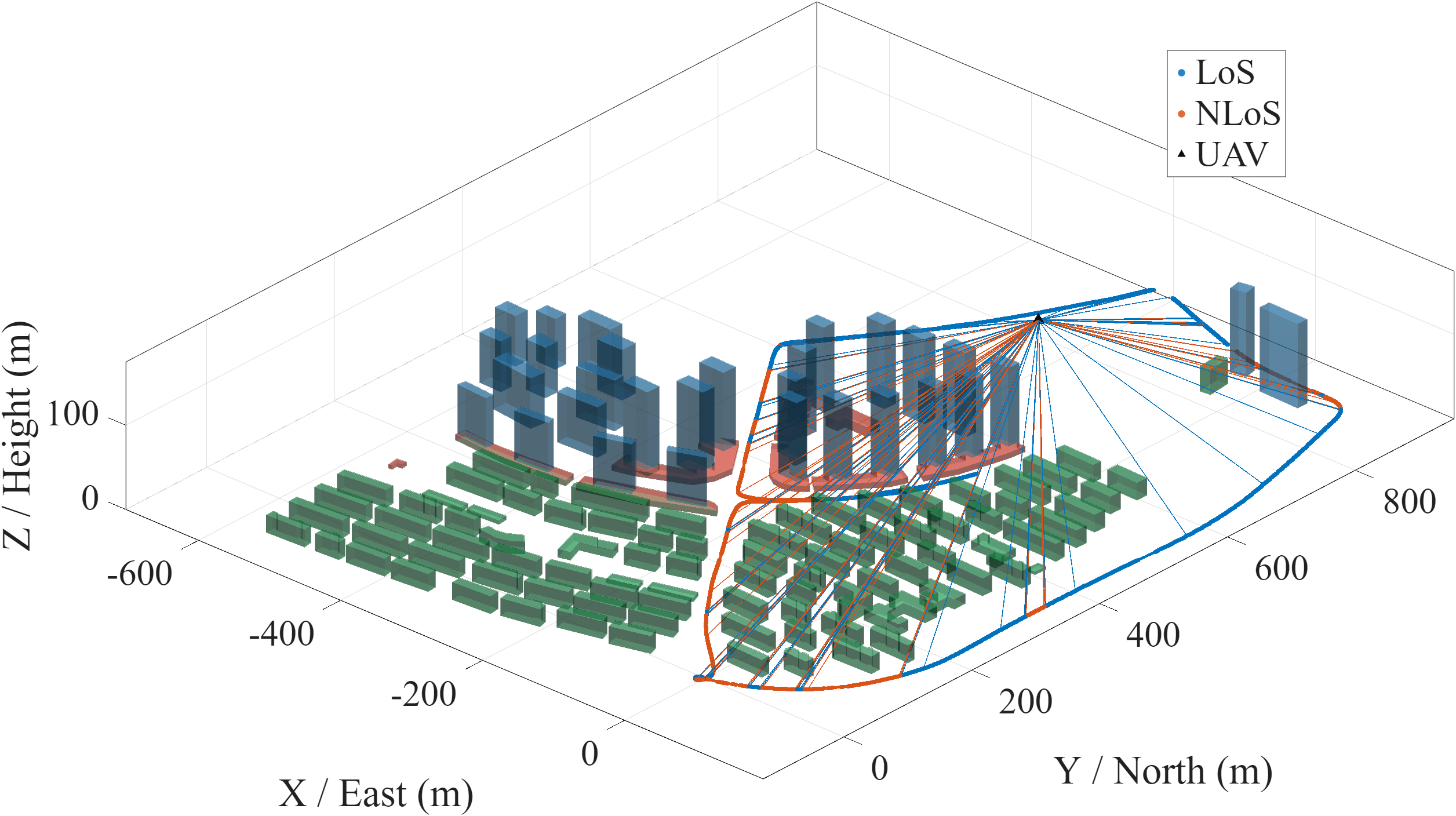}
	\caption{Reconstructed 3D urban scenario with identified LoS/NLoS states along the 2.85~GHz measurement trajectory.}
	\label{fig:sim_envir}
\end{figure}

Fig.~\ref{fig:sim_envir} shows the reconstructed three dimensional (3D) urban environment of the measurement area. To evaluate the Fresnel clearance margin of the Tx-Rx link based on geometric information, each building surface in the environment is represented by a mesh model composed of triangular facets. The mesh model is denoted as
\begin{equation}
	\mathcal{M}_{\mathrm{obs}}=\{\Delta_j\}_{j=1}^{J},
	\label{eq:mesh_model}
\end{equation}
where \(\mathcal{M}_{\mathrm{obs}}\) denotes the 3D environmental structural mesh, \(\Delta_j\) denotes the \(j\)-th triangular facet, and \(J\) is the total number of triangular facets. For \(N\) snapshots collected during the measurement, the transceiver positions can be expressed as
\begin{equation}
	\begin{aligned}
		\mathbf{T} &=[\mathbf{t}_1,\mathbf{t}_2,\ldots,\mathbf{t}_N]^{\mathrm{T}}\in\mathbb{R}^{N\times 3},\\
		\mathbf{R} &=[\mathbf{r}_1,\mathbf{r}_2,\ldots,\mathbf{r}_N]^{\mathrm{T}}\in\mathbb{R}^{N\times 3}.
	\end{aligned}
	\label{eq:tx_rx_matrix}
\end{equation}
where \(\mathbf{t}_i\in\mathbb{R}^3\) and \(\mathbf{r}_i\in\mathbb{R}^3\) denote the Tx and Rx coordinates of the \(i\)-th snapshot, respectively. For the \(i\)-th snapshot, the Tx-Rx link can be represented as
\begin{equation}
	\mathbf{l}_i(s)=\mathbf{r}_i+s(\mathbf{t}_i-\mathbf{r}_i),
	\label{eq:link_line}
\end{equation}
where \(\mathbf{l}_i(s)\) denotes a point on the \(i\)-th Tx-Rx link, and \(s\) is the normalized link position parameter with \(s\in[0,1]\). Along each Tx-Rx link, \(K\) sampling points are uniformly selected. The normalized position of the \(k\)-th sampling point is \(s_k=k/(K+1)\), and its 3D coordinate is \(\mathbf{p}_{i,k}=\mathbf{l}_i(s_k)\). Let the distances from this sampling point to the Tx and Rx be \(d^{\mathrm{Tx}}_{i,k}=\|\mathbf{p}_{i,k}-\mathbf{t}_i\|_2\) and \(d^{\mathrm{Rx}}_{i,k}=\|\mathbf{p}_{i,k}-\mathbf{r}_i\|_2\), respectively. The corresponding first Fresnel zone radius can be calculated as
\begin{equation}
	R_{F,i,k}
	=
	\sqrt{
		\frac{\lambda d^{\mathrm{Tx}}_{i,k}d^{\mathrm{Rx}}_{i,k}}
		{d^{\mathrm{Tx}}_{i,k}+d^{\mathrm{Rx}}_{i,k}}
	},
	\label{eq:fresnel_radius}
\end{equation}
where \(\lambda=c/f_c\) is the carrier wavelength, \(c\) is the speed of light, and \(f_c\) is the carrier frequency.

Conventional clearance margin evaluation is usually based on a two dimensional vertical path profile of the Tx-Rx link and calculates the height difference between the link centerline and the obstacle profile. However, in A2G communications, link blockage does not necessarily originate from obstacles directly below the link. It may also be caused by building edges intruding into the Fresnel zone.  Therefore, this paper evaluates the clearance margin by characterizing the 3D spatial relationship between link sampling points and surrounding obstruction surfaces. Specifically, the shortest Euclidean distance \(d_{i,k,j}\) from the sampling point \(\mathbf{p}_{i,k}\) to the \(j\)-th triangular facet \(\Delta_j\) is defined as
\begin{equation}
	d_{i,k,j}
	=
	\min_{\mathbf{q}\in \Delta_j}
	\|\mathbf{p}_{i,k}-\mathbf{q}\|_2 ,
	\label{eq:point_triangle_distance}
\end{equation}
where \(\mathbf{q}\) is an arbitrary point on \(\Delta_j\). To distinguish whether each sampling point is located inside or outside a building, we introduce a sign factor \(\xi_{i,k}\), which is assigned $+1$ if \(\mathbf{p}_{i,k}\) is outside the building and $-1$ if \(\mathbf{p}_{i,k}\) is inside the building. With this sign convention, the minimum signed distance from the sampling point to all building surfaces, denoted by \(D_s(\mathbf{p}_{i,k},\mathcal{M}_{\mathrm{obs}})\), is defined as
\begin{equation}
	D_s(\mathbf{p}_{i,k},\mathcal{M}_{\mathrm{obs}})
	=
	\xi_{i,k}\min_{1\le j\le J}d_{i,k,j}.
	\label{eq:signed_distance}
\end{equation}

Accordingly, the Fresnel clearance margin at the \(k\)-th sampling point of the \(i\)-th link is defined as
\begin{equation}
	m_{i,k}
	=
	D_s(\mathbf{p}_{i,k},\mathcal{M}_{\mathrm{obs}})-\eta R_{F,i,k}.
	\label{eq:sample_clearance}
\end{equation}
where \(\eta=0.6\) is the effective first Fresnel zone clearance ratio. When \(m_{i,k}>0\), no building intrusion occurs in the communication link at this sampling point. When \(m_{i,k}<0\), the nearest obstacle falls within the local Fresnel region, indicating a higher blockage risk. Since the propagation state of a Tx-Rx link is usually determined by the most unfavorable position along the path, the minimum Fresnel clearance margin of the \(i\)-th snapshot is defined as
\begin{equation}
	m_i=\min_{1\le k\le K}m_{i,k}.
	\label{eq:min_clearance}
\end{equation}

Finally, the minimum clearance margin is mapped to the geometric LoS prior probability through a sigmoid function
\begin{equation}
	p_i^{\mathrm{geo}}
	=
	\frac{1}{1+\exp(-m_i/s_g)} ,
	\label{eq:geo_prior}
\end{equation}
where \(p_i^{\mathrm{geo}}\) denotes the geometric LoS prior probability of the \(i\)-th snapshot, and \(s_g\) is a geometric transition scale used to make the clearance margin dimensionless before the sigmoid mapping.

\subsection{Channel-Feature-Aided Probability Fusion}

The LoS prior probability derived from geometric information provides a physically interpretable preliminary assessment of link blockage. However, the reconstructed environmental model inevitably suffers from limited geometric precision and model simplifications compared with the real-world propagation environment. As a result, the geometric prior may still be uncertain, especially in the transition regions between LoS and NLoS conditions. Since A2G channel characteristics usually exhibit distinct behaviors under different propagation states, the following channel features are further introduced as complementary evidence to refine the geometry-based prior for propagation state identification.

\noindent\textit{1) Strongest MPC Power:}
The strongest MPC power characterizes the dominant power component in the A2G channel. Let \(P_{i,l}=|h_i(\tau_{l})|^2\) denote the power of the \(l\)-th effective MPC in the \(i\)-th snapshot, and \(P_{\max,i}=\max_{1\le l\le L}P_{i,l}\). Its dB-domain value is given by \(P_{\max,i}^{\mathrm{dB}}=10\log_{10}\left(P_{\max,i}\right)\).

\noindent\textit{2) Excess Path Loss:}
Path loss characterizes the power attenuation of wireless signals. Based on the measured CIR, it can be expressed as
\begin{equation}
	PL_i[\mathrm{dB}]
	=
	-10\log_{10}
	\left(
	\frac{1}{|\mathcal{W}_i|}
	\sum_{u\in\mathcal{W}_i}
	\sum_{l=1}^{L}
	|h_u(\tau_l)|^2
	\right),
	\label{eq:stage2_path_loss}
\end{equation}
where \(\mathcal{W}_i\) is the sliding window centered at the \(i\)-th snapshot and used to smooth SSF. The free space path loss (FSPL) is given by
\begin{equation}
	FSPL_i[\mathrm{dB}]
	=
	20\log_{10}
	\left(
	\frac{4\pi d_i f_c}{c}
	\right),
	\label{eq:stage2_fspl}
\end{equation}
where \(d_i\) is the transceiver distance of the \(i\)-th snapshot. Accordingly, the excess PL relative to FSPL is defined as
\begin{equation}
	\Delta PL_i[\mathrm{dB}]=PL_i-FSPL_i.
	\label{eq:stage2_delta_pl}
\end{equation}

\noindent\textit{3) RMS Delay Spread:}
The RMS delay spread characterizes the delay dispersion of MPCs and is defined as
\begin{equation}
	\tau_{\mathrm{rms},i}
	=
	\sqrt{
		\frac{\sum_{l=1}^{L}P_{i,l}\tau_{i,l}^2}{\sum_{l=1}^{L}P_{i,l}}
		-
		\left(
		\frac{\sum_{l=1}^{L}P_{i,l}\tau_{i,l}}{\sum_{l=1}^{L}P_{i,l}}
		\right)^2
	}.
	\label{eq:stage2_rms_ds}
\end{equation}

\noindent\textit{4) Rician K-factor:}
The Rician K-factor measures the power ratio between the dominant propagation component and the remaining MPCs, and is expressed in the dB domain as
\begin{equation}
	KF_i[\mathrm{dB}]
	=
	10\log_{10}
	\left(
	\frac{P_{\max,i}}
	{\sum_{l=1}^{L}P_{i,l}-P_{\max,i}}
	\right).
	\label{eq:stage2_k_factor}
\end{equation}

\noindent\textit{5) Skewness of Received Amplitude:}
Skewness describes the asymmetry of the MPC amplitude distribution, is defined as
\begin{equation}
	S_i
	=
	\frac{
		\frac{1}{L}\sum_{l=1}^{L}
		\left(|h_i(\tau_{i,l})|-\mu_{h,i}\right)^3
	}
	{\sigma_{h,i}^3}.
	\label{eq:stage2_skewness}
\end{equation}
where \(\mu_{h,i}\) and \(\sigma_{h,i}\) are the mean and standard deviation of the MPC amplitudes, respectively.

\noindent\textit{6) Kurtosis of Received Amplitude:}
Kurtosis describes the peakedness of the MPC amplitude distribution and is expressed as
\begin{equation}
	\kappa_i
	=
	\frac{
		\frac{1}{L}\sum_{l=1}^{L}
		\left(|h_i(\tau_{i,l})|-\mu_{h,i}\right)^4
	}
	{\sigma_{h,i}^4}.
	\label{eq:stage2_kurtosis}
\end{equation}

In summary, the channel feature vector of the \(i\)-th snapshot is expressed as
\begin{equation}
	\mathbf{x}_i
	=
	\left[
	P_{\max,i}^{\mathrm{dB}},
	\Delta PL_i,
	\tau_{\mathrm{rms},i},
	KF_i,
	S_i,
	\kappa_i
	\right]^{\mathrm{T}}.
	\label{eq:stage2_feature_vector}
\end{equation}

In the pseudo label generation stage, the goal is not to label all snapshots, but to provide high confidence pseudo labeled samples for weakly supervised random forest (RF) training. Specifically, a pseudo label is assigned only when the geometric prior and channel feature evidence consistently support the same propagation state direction. For the geometry prior, if \(p_i^{\mathrm{geo}}>0.9\), the \(i\)-th snapshot is considered to have high LoS geometric confidence, and if \(p_i^{\mathrm{geo}}<0.1\), it is considered to have high NLoS geometric confidence. For the channel features, the lower and upper quartiles of each feature are calculated as decision thresholds. Let \(Q_{1,\cdot}\) and \(Q_{3,\cdot}\) denote the first and third quartiles of the corresponding feature. The numbers of LoS supporting and NLoS supporting channel feature votes are defined as
\begin{equation}
	\begin{aligned}
		v_i^{\mathrm{L}}={}&
		I\!\left(P_{\max,i}^{\mathrm{dB}}>Q_{3,P_{\max}}\right)
		+I\!\left(\Delta PL_i<Q_{1,\Delta PL}\right)\\
		&+I\!\left(\tau_{\mathrm{rms},i}<Q_{1,\tau}\right)
		+I\!\left(KF_i>Q_{3,KF}\right)\\
		&+I\!\left(S_i>Q_{3,S}\right)
		+I\!\left(\kappa_i>Q_{3,\kappa}\right),
	\end{aligned}
	\label{eq:stage2_los_vote}
\end{equation}
\begin{equation}
	\begin{aligned}
		v_i^{\mathrm{N}}={}&
		I\!\left(P_{\max,i}^{\mathrm{dB}}<Q_{1,P_{\max}}\right)
		+I\!\left(\Delta PL_i>Q_{3,\Delta PL}\right)\\
		&+I\!\left(\tau_{\mathrm{rms},i}>Q_{3,\tau}\right)
		+I\!\left(KF_i<Q_{1,KF}\right)\\
		&+I\!\left(S_i<Q_{1,S}\right)
		+I\!\left(\kappa_i<Q_{1,\kappa}\right),
	\end{aligned}
	\label{eq:stage2_nlos_vote}
\end{equation}
where \(I(\cdot)\) is the indicator function. High values of \(P_{\max,i}^{\mathrm{dB}}\), \(KF_i\), \(S_i\), and \(\kappa_i\), together with low values of \(\Delta PL_i\) and \(\tau_{\mathrm{rms},i}\), support the LoS state; the opposite conditions support the NLoS state. Features that do not satisfy the corresponding quartile criteria do not contribute a vote. Let \(M\) denote the number of available channel features used for voting. The voting threshold is set as \(N_v=\lceil(M+1)/2\rceil\), requiring more than half of the available channel features to support the same propagation state. Therefore, the pseudo label can be expressed as
\begin{equation}
	\tilde{z}_i
	=
	\begin{cases}
		1,  & p_i^{\mathrm{geo}}>0.9,\ v_i^{\mathrm{L}}\ge N_v,\\
		0,  & p_i^{\mathrm{geo}}<0.1,\ v_i^{\mathrm{N}}\ge N_v,\\
		-1, & \mathrm{otherwise},
	\end{cases}
	\label{eq:stage2_pseudo_label}
\end{equation}
where \(\tilde{z}_i\in\{-1,0,1\}\) denotes the pseudo label of the \(i\)-th snapshot, with \(\tilde{z}_i=1\) and \(\tilde{z}_i=0\) representing high confidence LoS and high confidence NLoS, respectively, and \(\tilde{z}_i=-1\) representing an unlabeled sample. The high confidence pseudo labeled samples are used to train an RF classifier in the channel feature space. Given the feature vector \(\mathbf{x}_i\), the classifier outputs a channel feature based LoS probability estimate \(p_i^{\mathrm{radio}}\). Since the classifier is trained with pseudo labels, \(p_i^{\mathrm{radio}}\) is interpreted as a probability estimate rather than a calibrated posterior probability.

Finally, the channel feature based LoS probability estimate and the geometric LoS prior are fused in the logit domain
\begin{equation}
	\mathrm{logit}\!\left(p_i^{\mathrm{fused}}\right)
	=
	\alpha\,\mathrm{logit}\!\left(p_i^{\mathrm{radio}}\right)
	+
	\beta\,\mathrm{logit}\!\left(p_i^{\mathrm{geo}}\right),
	\label{eq:stage2_logit_fusion}
\end{equation}
where \(\alpha\) and \(\beta\) are weighting factors that control the relative contributions of the channel feature evidence and geometric prior, respectively. The corresponding fused LoS probability is given by
\begin{equation}
	p_i^{\mathrm{fused}}
	=
	\frac{1}
	{1+\exp\left[
		-\alpha\,\mathrm{logit}\!\left(p_i^{\mathrm{radio}}\right)
		-\beta\,\mathrm{logit}\!\left(p_i^{\mathrm{geo}}\right)
		\right]}.
	\label{eq:stage2_fused_prob}
\end{equation}

\subsection{HMM-Based Spatial Consistency Smoothing}

The fused LoS probabilities obtained through the above procedure are estimated independently for individual snapshots and may therefore contain isolated fluctuations. However, in the considered A2G measurement scenario, as the ground vehicle moves continuously along the route, the propagation state usually remains stable over a certain spatial distance. Therefore, a hidden Markov model is introduced to infer a spatially consistent LoS/NLoS state sequence from the fused probability sequence. Let \(z_i\in\{0,1\}\) be the hidden propagation state of the \(i\)-th snapshot, where \(z_i=1\) denotes LoS, and \(z_i=0\) denotes NLoS. The fused LoS probability \(p_i^{\mathrm{fused}}\) is used to construct the HMM emission scores
\begin{equation}
	b_i(1)=p_i^{\mathrm{fused}},
	\qquad
	b_i(0)=1-p_i^{\mathrm{fused}},
	\label{eq:stage3_emission_score}
\end{equation}
where \(b_i(z_i)\) denotes the emission score for state \(z_i\). To avoid manually assigning fixed transition probabilities, an initial hard decision sequence is first obtained as
\begin{equation}
	z_i^{(0)}
	=
	\begin{cases}
		1, & p_i^{\mathrm{fused}}\ge 0.5,\\
		0, & p_i^{\mathrm{fused}}<0.5.
	\end{cases}
	\label{eq:stage3_initial_state}
\end{equation}

Based on this initial sequence, the average spatial persistence lengths of contiguous LoS and NLoS segments are calculated and denoted by \(\bar{L}_{\mathrm{LoS}}\) and \(\bar{L}_{\mathrm{NLoS}}\), respectively. The length of each segment is computed from the cumulative Rx displacement. The directional spatial transition rate parameters are then approximated as
\begin{equation}
	\lambda_{\mathrm{L}\rightarrow\mathrm{N}}
	=
	\frac{1}{\bar{L}_{\mathrm{LoS}}},
	\qquad
	\lambda_{\mathrm{N}\rightarrow\mathrm{L}}
	=
	\frac{1}{\bar{L}_{\mathrm{NLoS}}}.
	\label{eq:stage3_directional_lambda}
\end{equation}
where \(\lambda_{\mathrm{L}\rightarrow\mathrm{N}}\) and \(\lambda_{\mathrm{N}\rightarrow\mathrm{L}}\) denote the spatial transition rate parameters for leaving the LoS and NLoS states, respectively. For the Rx displacement \(\Delta s_i\) between the \((i-1)\)-th and \(i\)-th snapshots, the directional switching probabilities are defined as
\begin{equation}
	p_{\mathrm{L}\rightarrow\mathrm{N}}(\Delta s_i)
	=
	1-\exp(-\lambda_{\mathrm{L}\rightarrow\mathrm{N}}\Delta s_i),
	\label{eq:stage3_switch_ln}
\end{equation}
\begin{equation}
	p_{\mathrm{N}\rightarrow\mathrm{L}}(\Delta s_i)
	=
	1-\exp(-\lambda_{\mathrm{N}\rightarrow\mathrm{L}}\Delta s_i).
	\label{eq:stage3_switch_nl}
\end{equation}

With the state order \([\mathrm{NLoS}, \mathrm{LoS}]\), the transition matrix from \(z_{i-1}\) to \(z_i\) is
\begin{equation}
	\mathbf{A}_i
	=
	\begin{bmatrix}
		1-p_{\mathrm{N}\rightarrow\mathrm{L}}(\Delta s_i) &
		p_{\mathrm{N}\rightarrow\mathrm{L}}(\Delta s_i)\\
		p_{\mathrm{L}\rightarrow\mathrm{N}}(\Delta s_i) &
		1-p_{\mathrm{L}\rightarrow\mathrm{N}}(\Delta s_i)
	\end{bmatrix}.
	\label{eq:stage3_transition_matrix}
\end{equation}

Finally, Viterbi decoding is performed in the log domain using the emission scores and transition matrices to obtain the final spatially smoothed state sequence \(\hat{\mathbf{z}}\).

\subsection{Identification Accuracy Evaluation}

To quantitatively evaluate the proposed propagation state identification method, the propagation state of each snapshot is manually annotated using synchronized videos recorded from the receiver vehicle toward the UAV. The resulting annotations are regarded as ground-truth labels. It should be noted that these labels are used exclusively for performance evaluation and are not involved in pseudo label extraction, channel feature fusion, or HMM-based smoothing.

\begin{table}[t]
	\renewcommand{\arraystretch}{1.3}
	\caption{Number of high confidence pseudo labeled snapshots used for weakly supervised RF training.}
	\label{tab:pseudo_label_snapshots}
	\begin{center}
		\begin{tabularx}{\linewidth}{
				>{\centering\arraybackslash}X|
				>{\centering\arraybackslash}X|
				>{\centering\arraybackslash}X|
				>{\centering\arraybackslash}X|
				>{\centering\arraybackslash}X|
				>{\centering\arraybackslash}X}
			\hline \hline
			\textbf{Frequency} &
			\textbf{\makecell{Total\\Snapshot}} &
			\textbf{\makecell{Pseudo\\LoS}} &
			\textbf{\makecell{Pseudo\\NLoS}} &
			\textbf{Unlabeled} &
			\textbf{\makecell{Pseudo-\\label\\ratio}} \\
			\hline
			2.85~GHz & 2696 & 378 & 236 & 2082 & 22.8\% \\
			\hline
			4.6~GHz  & 2618 & 430 & 166 & 2022 & 22.8\% \\
			\hline
			7.25~GHz & 1378 & 174 & 28  & 1176 & 14.7\% \\
			\hline \hline
		\end{tabularx}
	\end{center}
\end{table}

Table~\ref{tab:pseudo_label_snapshots} summarizes the numbers of high confidence pseudo labeled snapshots selected according to the proposed decision criterion. At 2.85 and 4.6~GHz, 22.8\% of the snapshots are selected for weakly supervised RF training, while the remaining snapshots are unlabeled. At 7.25~GHz, the pseudo label ratio decreases to 14.7\%. This reduction is mainly attributed to the higher path loss and greater sensitivity to blockage induced attenuation at 7.25~GHz. Under blockage conditions, the received signal is more likely to approach the noise floor, significantly reducing both the number of effective snapshots and the high confidence samples available for pseudo labeling. This conservative selection strategy avoids introducing unreliable pseudo labels into RF training, especially for severely attenuated snapshots.

\begin{figure}[!t]
	\centering
	\includegraphics[width=1\columnwidth]{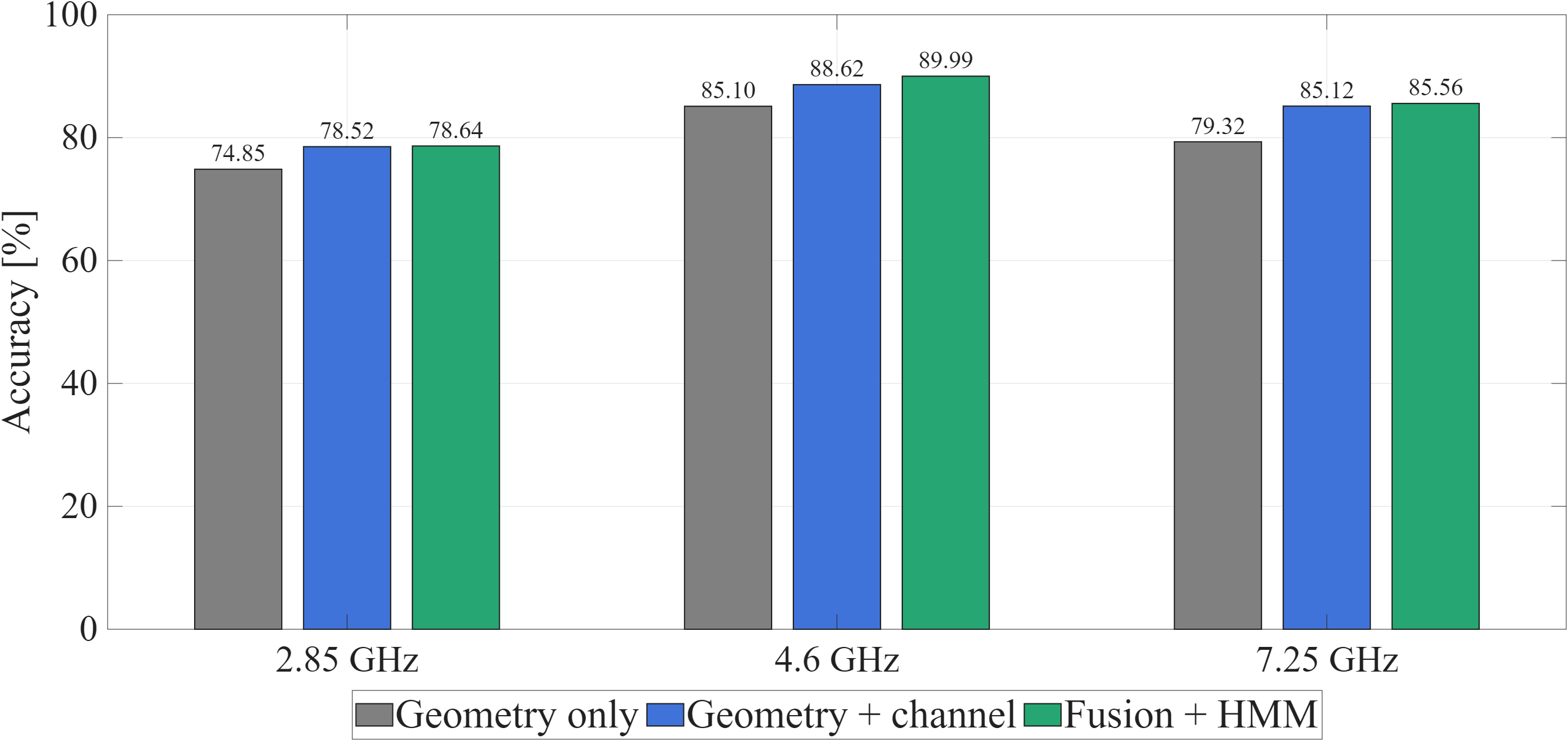}
	\caption{LoS/NLoS identification accuracy comparison among geometry-only decision, geometry-channel fusion, and HMM-based smoothing.}
	\label{fig:accuracy_comparison}
\end{figure}

The identification accuracy is evaluated by comparing the propagation state identified for each snapshot with its manually annotated ground-truth label. The accuracy is defined as the proportion of correctly identified LoS and NLoS snapshots among all evaluated snapshots, expressed as

\begin{equation}
	\mathrm{Accuracy}
	=
	\frac{TP+TN}{TP+TN+FP+FN},
	\label{eq:identification_accuracy}
\end{equation}
where \(TP\) denotes the number of LoS snapshots correctly identified as LoS, \(TN\) denotes the number of NLoS snapshots correctly identified as NLoS, \(FP\) denotes the number of NLoS snapshots incorrectly identified as LoS, and \(FN\) denotes the number of LoS snapshots incorrectly identified as NLoS. Therefore, a higher accuracy indicates that a larger proportion of snapshots are assigned the correct propagation state labels.

Fig.~\ref{fig:accuracy_comparison} compares the identification accuracies obtained using the geometry-only decision, geometry-and-channel fusion, and the HMM-smoothed result. Incorporating channel features increases the accuracies from 74.85\%, 85.10\%, and 79.32\% to 78.52\%, 88.62\%, and 85.12\% at 2.85, 4.6, and 7.25~GHz, respectively. This corresponds to improvements of 3.67, 3.52, and 5.80 percentage points over the geometry-only method, demonstrating that the channel features provide effective complementary information when the simplified geometric prior cannot fully characterize the actual propagation environment. After HMM smoothing, the final accuracies further increase to 78.64\%, 89.99\%, and 85.56\%, respectively. Although the additional accuracy improvements introduced by the HMM are relatively limited, its primary contribution is to suppress isolated state fluctuations and improve the spatial consistency of the identified state sequences. The final HMM-smoothed LoS/NLoS states at 2.85~GHz are further visualized along the measurement trajectory in Fig.~\ref{fig:sim_envir}.

\section{Multi-frequency Channel Characterization}

\subsection{Power Delay Profile}

\begin{figure*}[!t]
	\centering
	\subfloat[]{\includegraphics[width=0.68\columnwidth]{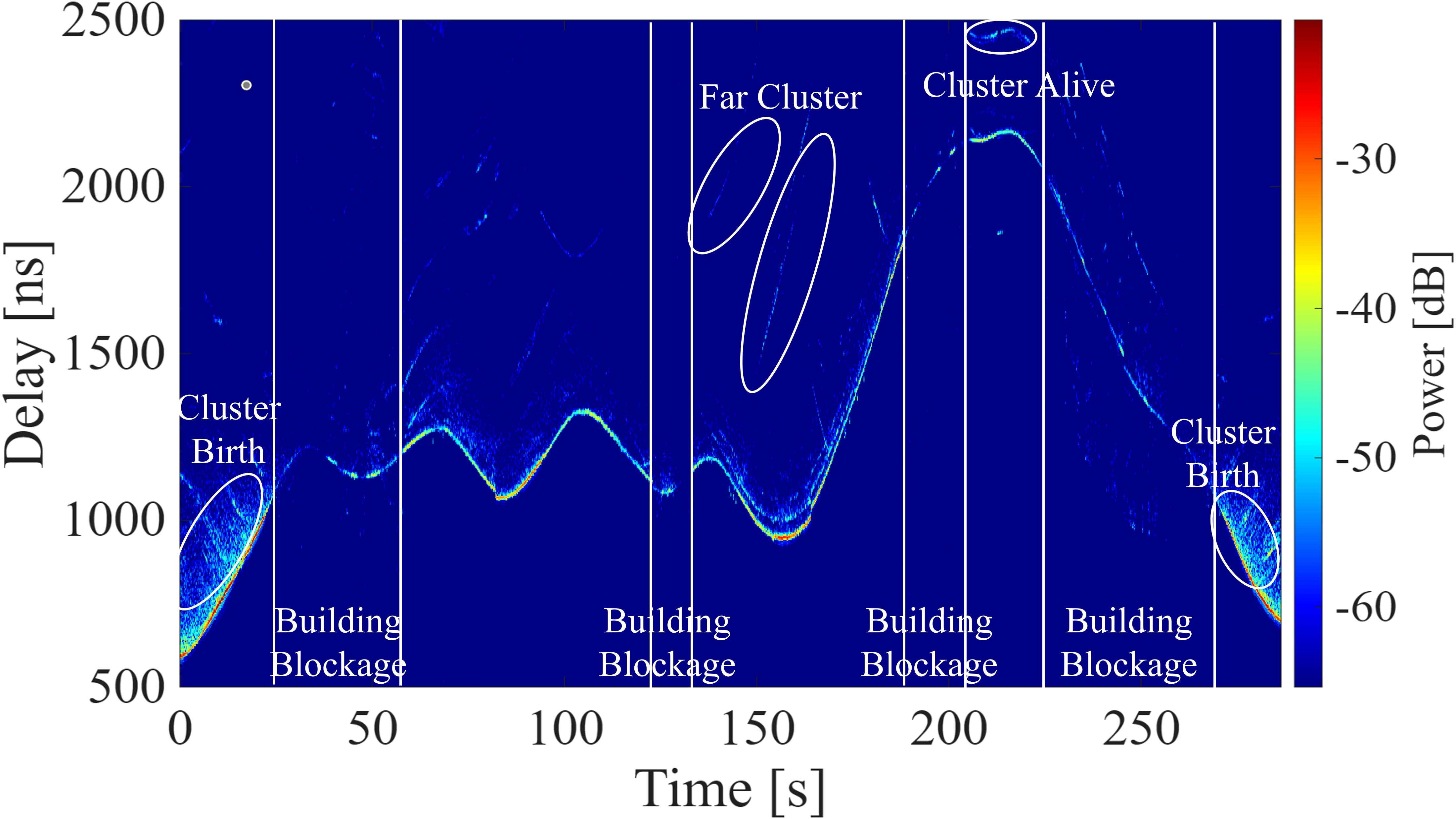}%
		\label{fig:PDP_2.85G}}
	\hfil
	\subfloat[]{\includegraphics[width=0.68\columnwidth]{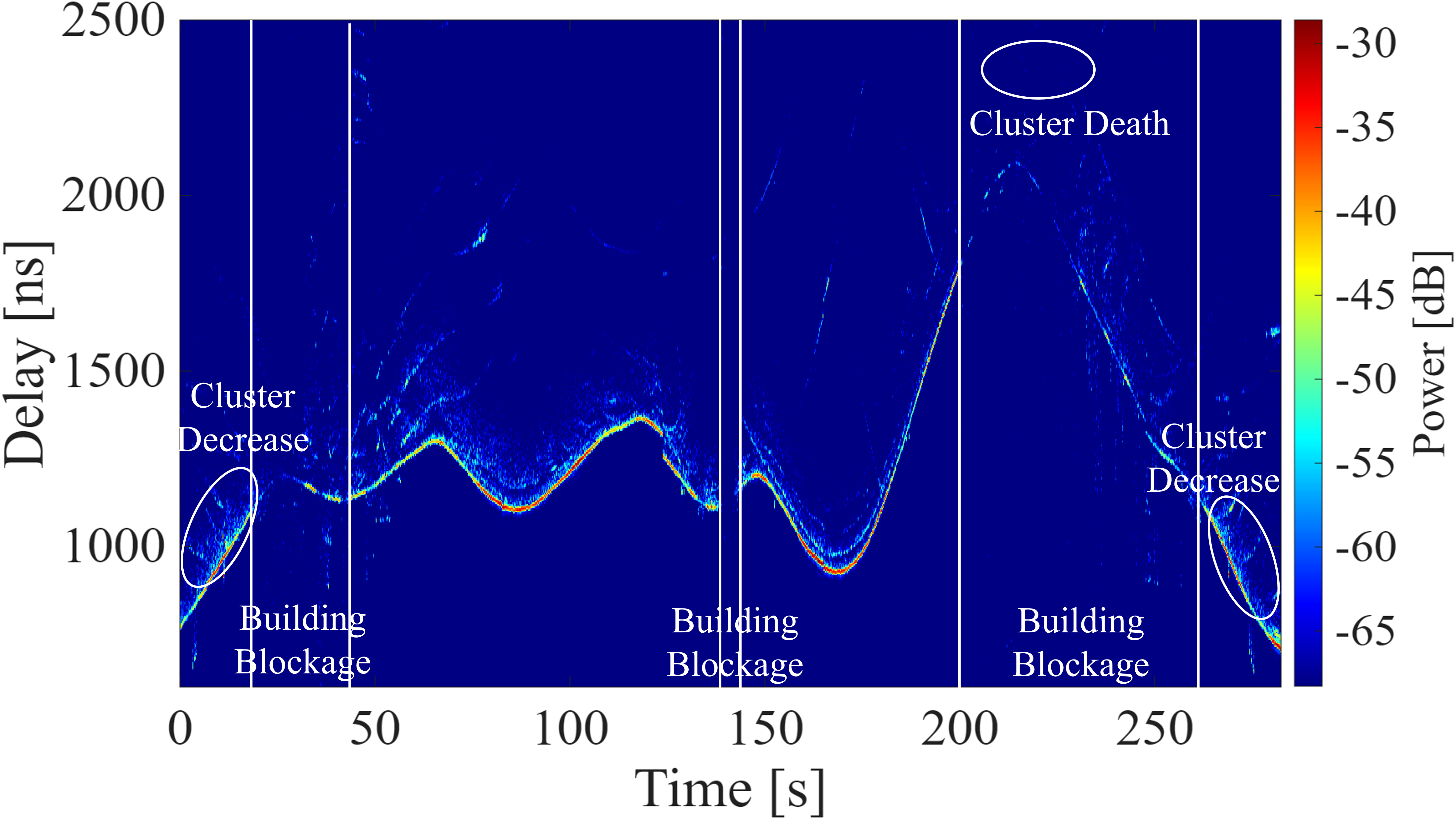}%
		\label{fig:PDP_4.6G}}
	\hfil
	\subfloat[]{\includegraphics[width=0.68\columnwidth]{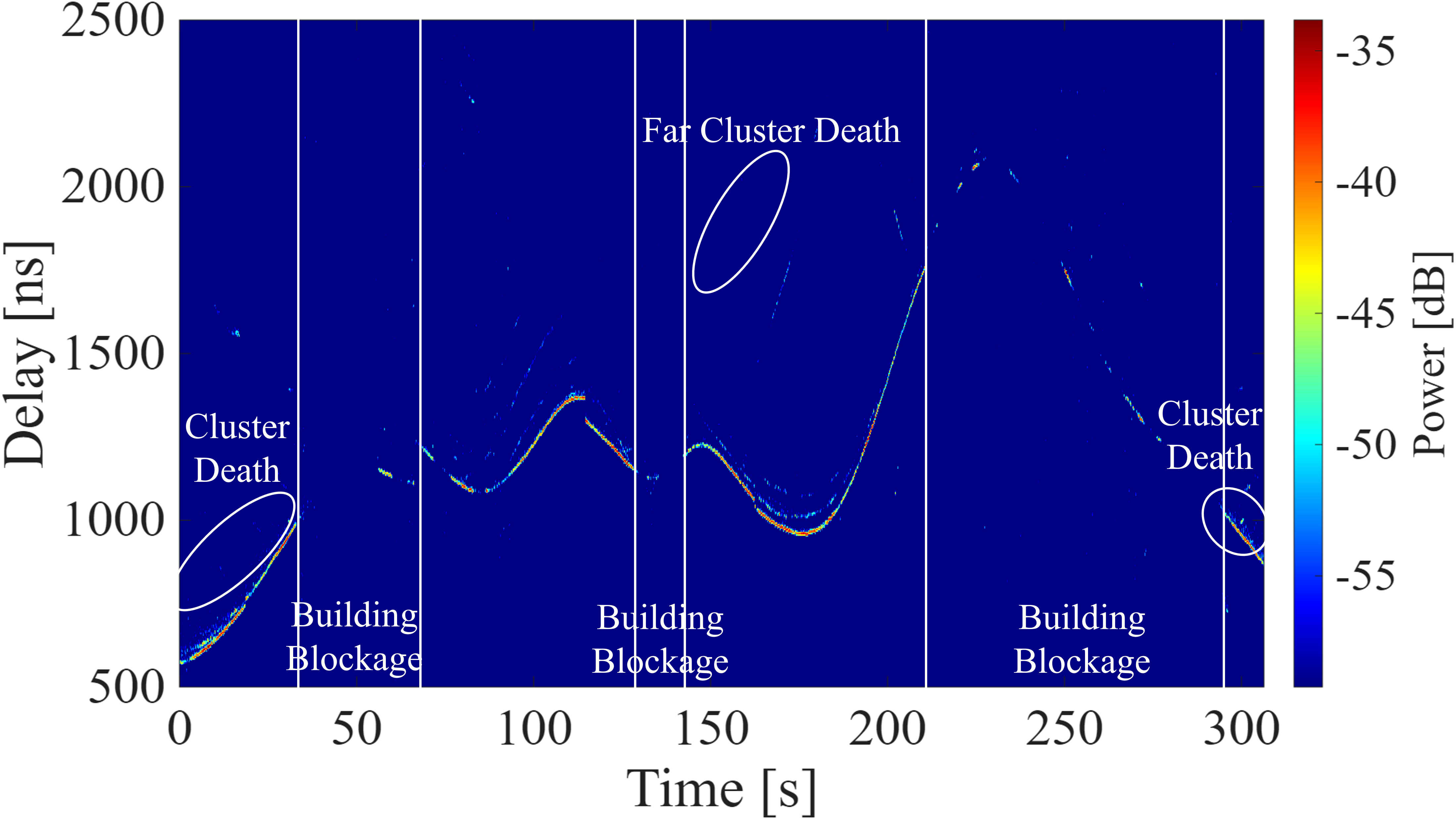}%
		\label{fig:PDP_7.25G}}
	\caption{Measured power delay profiles at different frequency bands: (a) 2.85~GHz; (b) 4.6~GHz; (c) 7.25~GHz.}
	\label{fig:PDP}
\end{figure*}

\begin{figure}[!t]
	\centering
	\subfloat[]{\includegraphics[width=0.49\columnwidth]{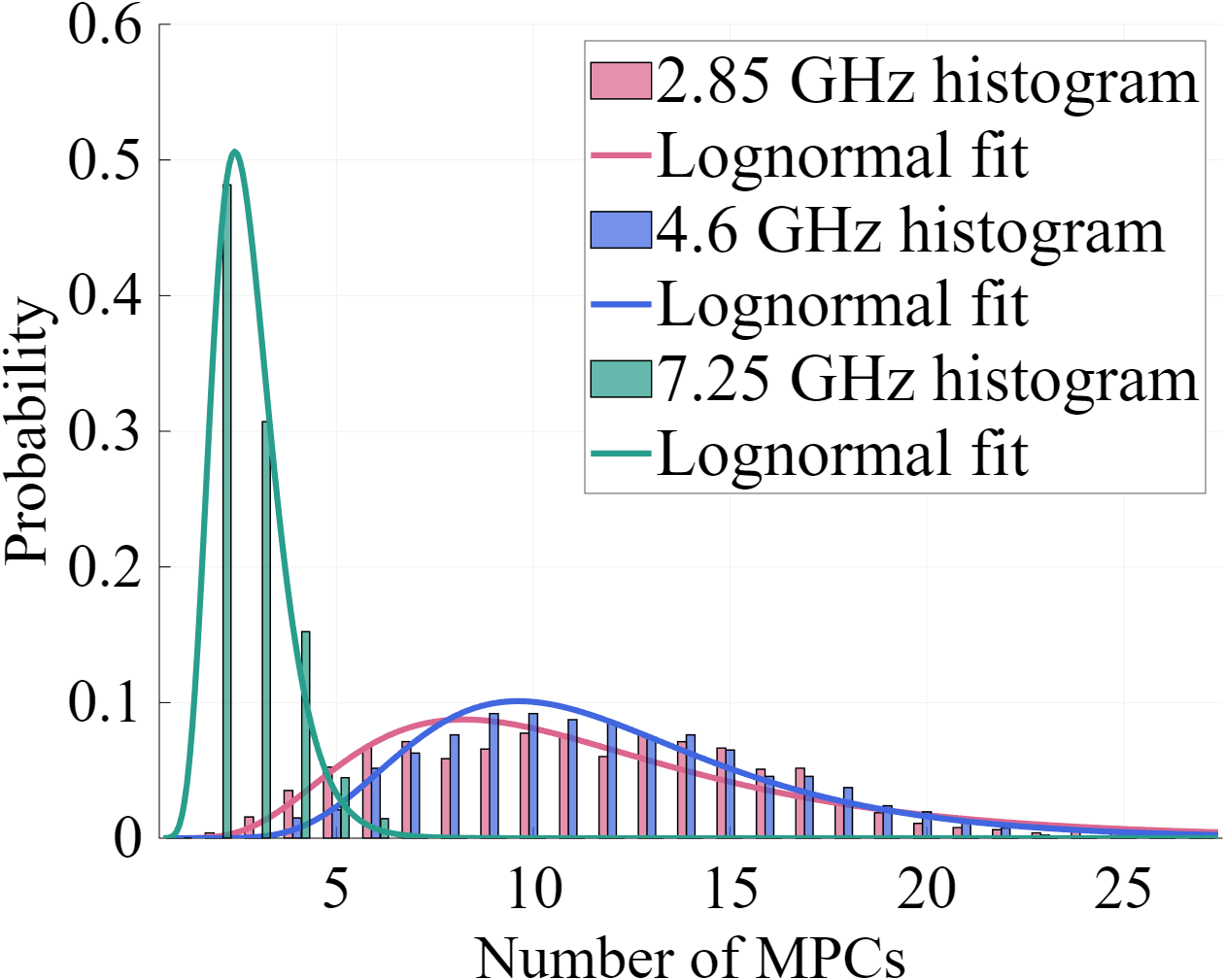}%
		\label{fig:mpc_num_los}}
	\hfil
	\subfloat[]{\includegraphics[width=0.49\columnwidth]{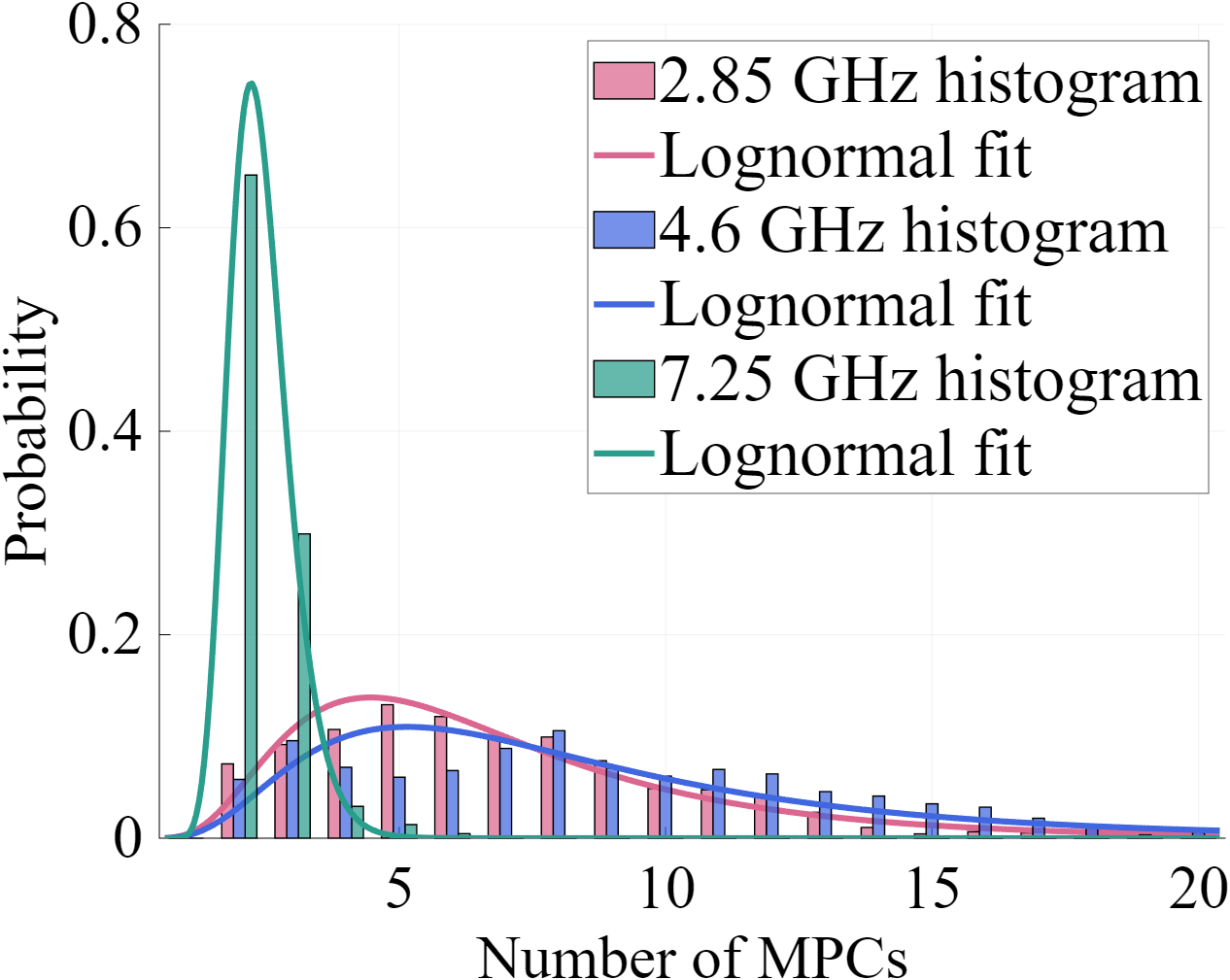}%
		\label{fig:mpc_num_nlos}}
	\caption{PDFs of the number of MPCs across different frequency bands: (a) LoS scenario; (b) NLoS scenario.}
	\label{fig:mpc_num}
\end{figure}

Fig.~\ref{fig:PDP}(a)--(c) show the measured PDPs at 2.85, 4.6, and 7.25~GHz, respectively. Since the three measurements were conducted along the same route, the PDPs exhibit similar variation trends. Regarding the power distribution, the peak powers of the strongest path at the three bands are -20.8, -28.5, and -33.8~dB, respectively. Meanwhile, the observable dynamic range decreases from about 45~dB to 25~dB. In addition, obvious attenuation or even disappearance of Rx-side local scattering clusters and far scattering clusters can be observed from the PDPs in the initial and final segments of the trajectory and during building-blockage intervals. This phenomenon becomes particularly pronounced at 7.25~GHz, where only a limited number of MPCs remain detectable under severe blockage conditions. Fig.~\ref{fig:mpc_num} further shows the distributions of the MPC number under LoS and NLoS conditions, and they can be well approximated by lognormal distributions. Overall, the number of MPCs under LoS conditions is larger than that under NLoS conditions. The median numbers of MPCs under LoS conditions are 11, 11, and 3 at 2.85, 4.6, and 7.25~GHz, respectively, while the corresponding values under NLoS conditions are 6, 8, and 2. The MPC number at 7.25~GHz is significantly lower than those at 2.85 and 4.6~GHz. This result indicates that the high-frequency A2G channel exhibits higher sparsity, mainly because higher propagation loss causes many weak MPCs to fall below the noise floor.

\subsection{Path Loss}

\begin{figure*}[!t]
	\centering
	\subfloat[]{\includegraphics[width=0.32\textwidth]{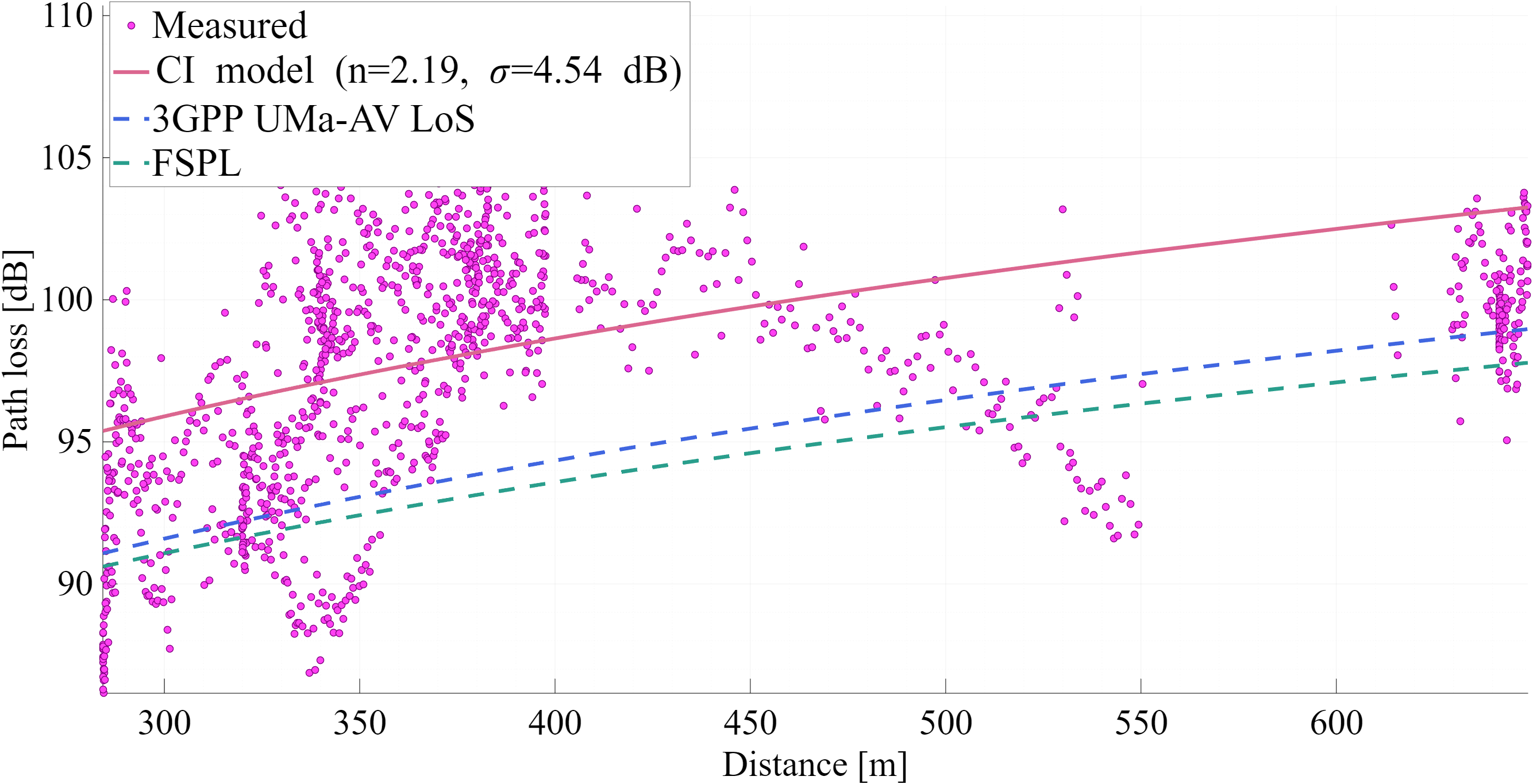}%
		\label{fig:pl_los_2850}}
	\hfil
	\subfloat[]{\includegraphics[width=0.32\textwidth]{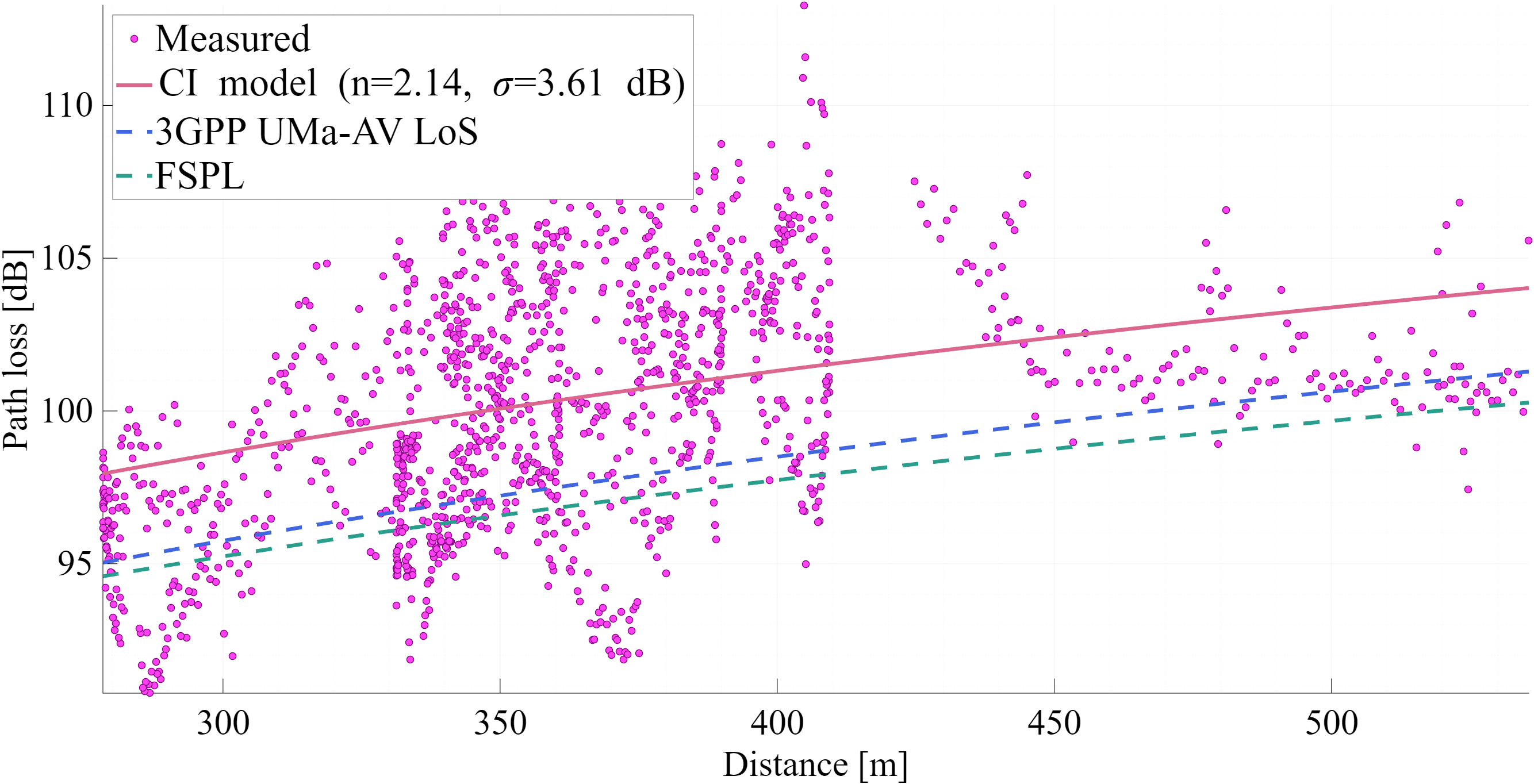}%
		\label{fig:pl_los_4600}}
	\hfil
	\subfloat[]{\includegraphics[width=0.32\textwidth]{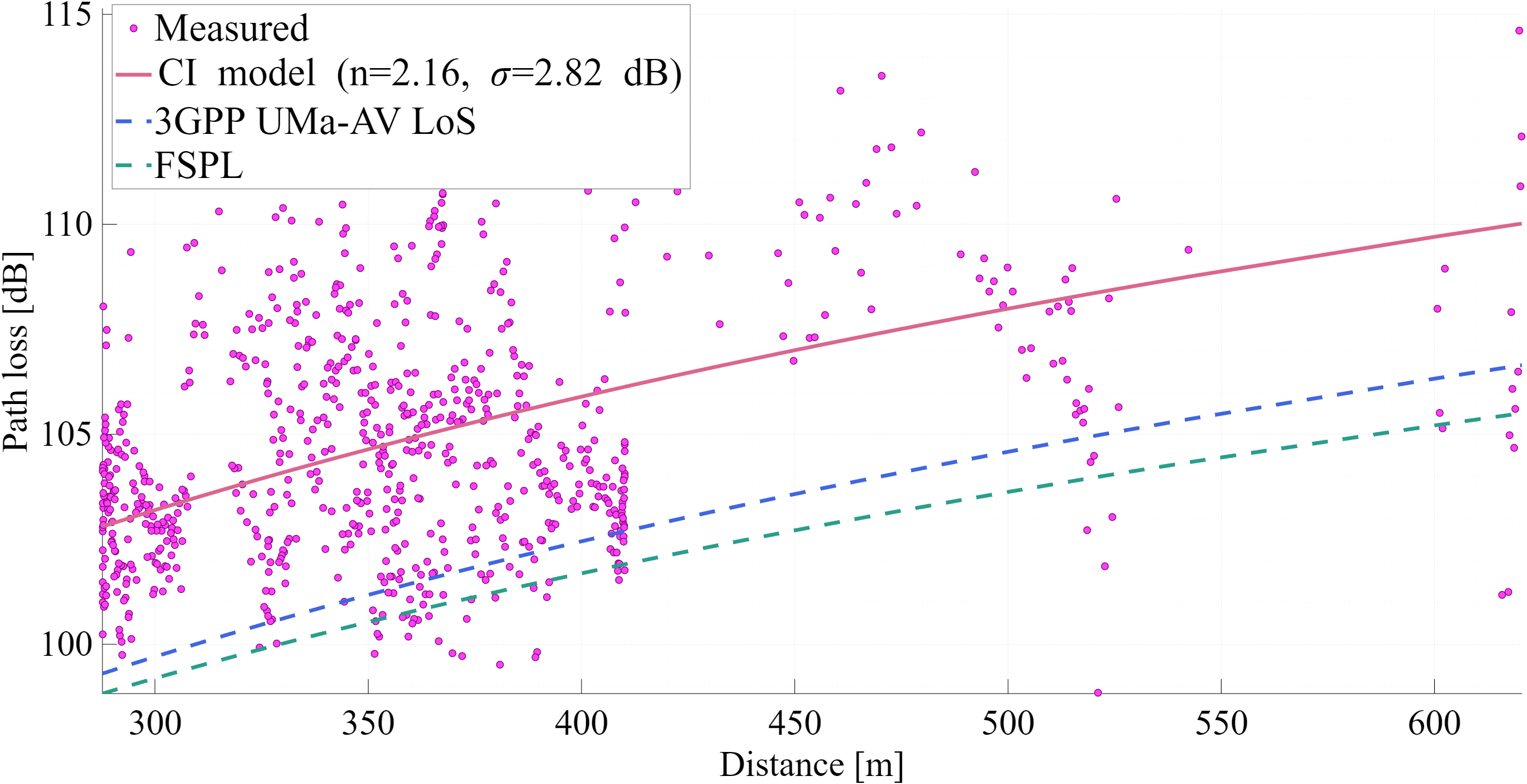}%
		\label{fig:pl_los_7250}}
	\\
	\subfloat[]{\includegraphics[width=0.32\textwidth]{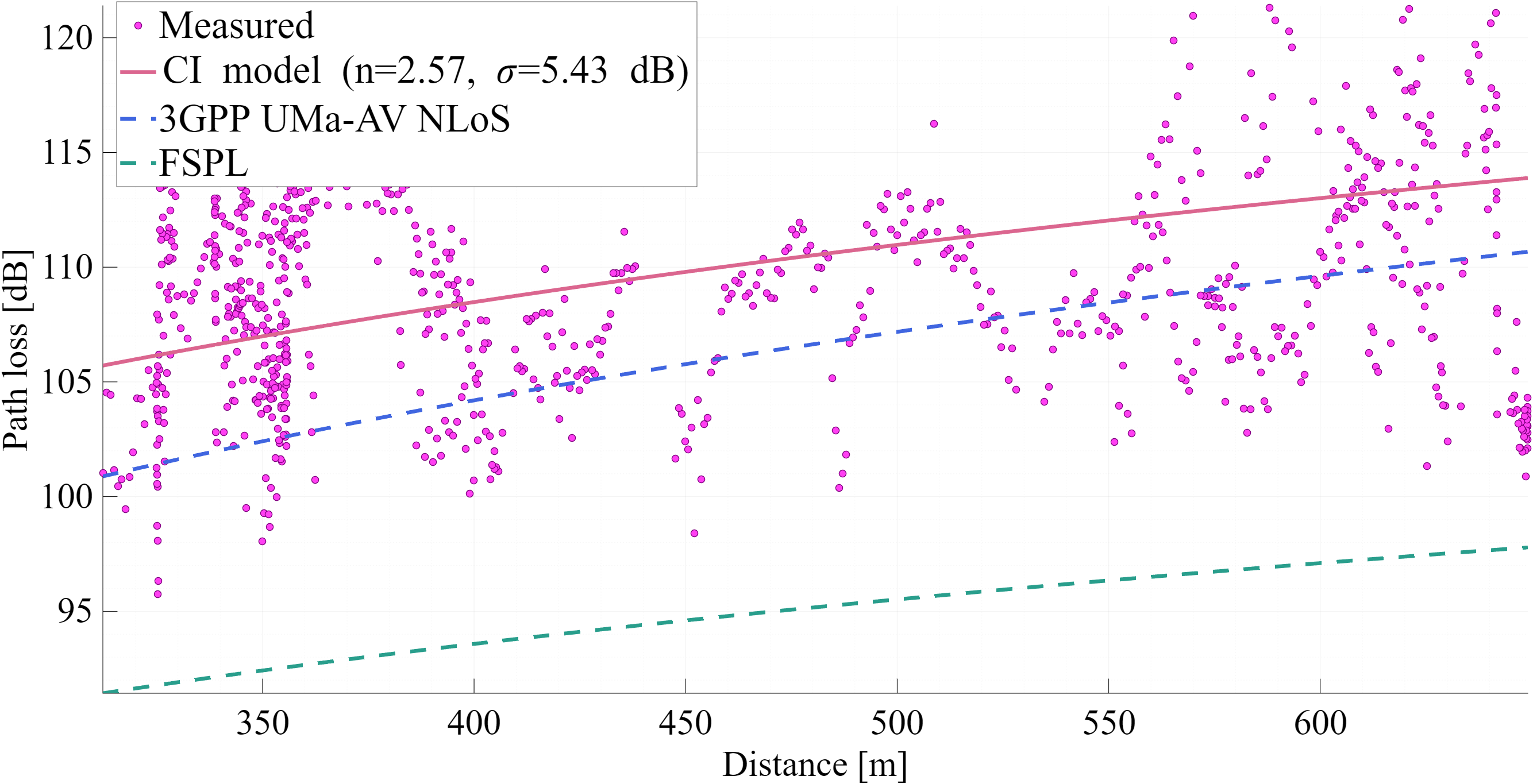}%
		\label{fig:pl_nlos_2850}}
	\hfil
	\subfloat[]{\includegraphics[width=0.32\textwidth]{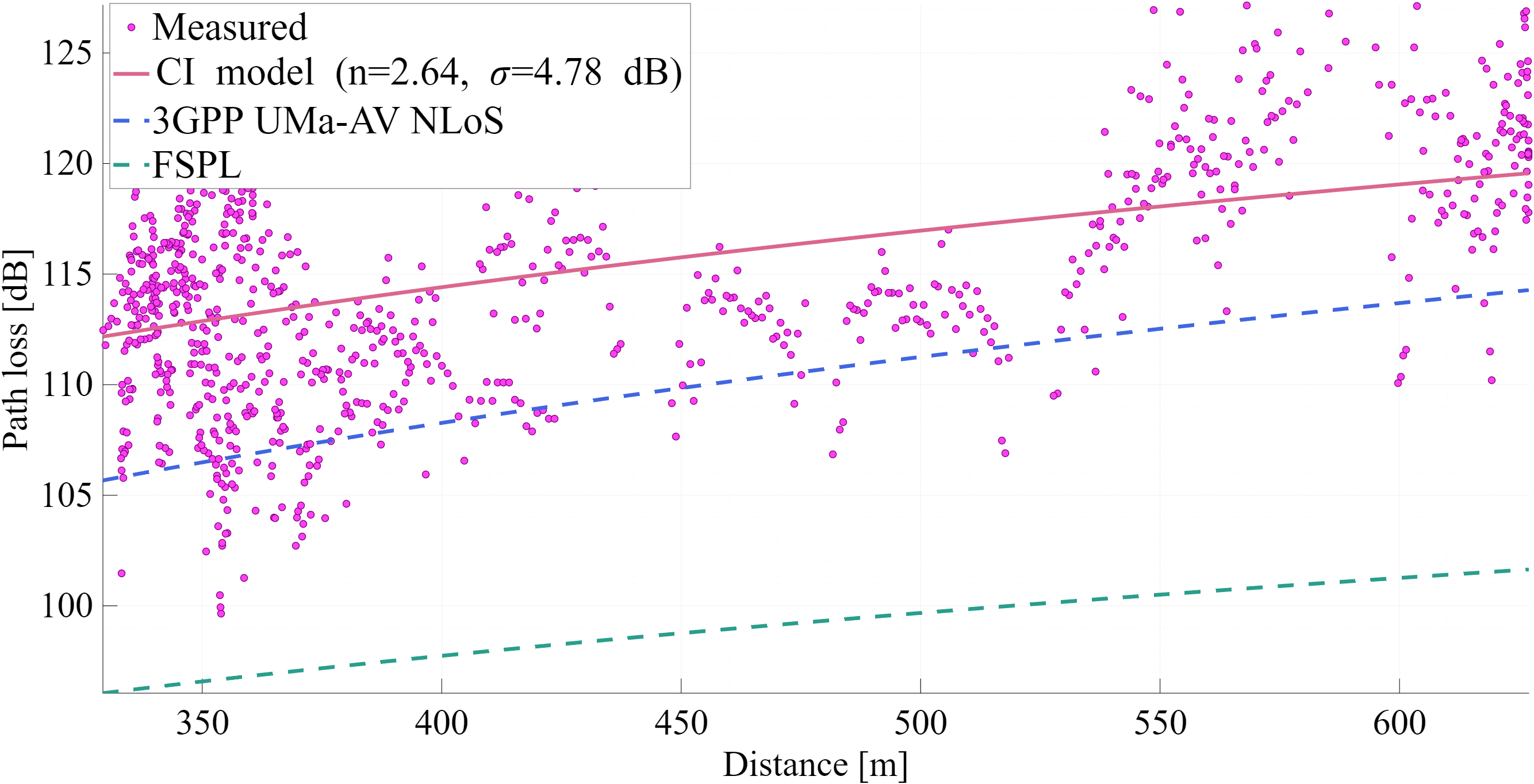}%
		\label{fig:pl_nlos_4600}}
	\hfil
	\subfloat[]{\includegraphics[width=0.32\textwidth]{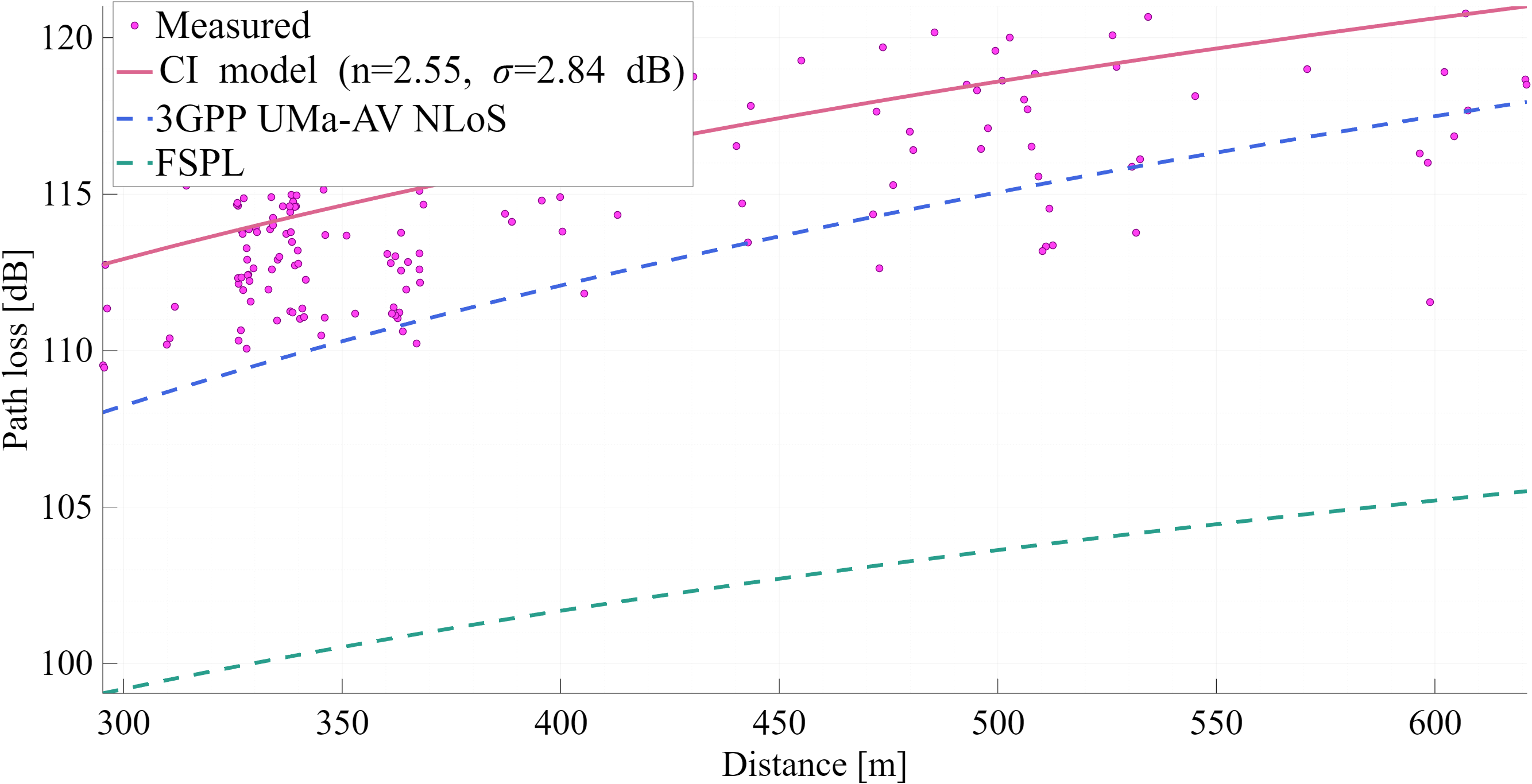}%
		\label{fig:pl_nlos_7250}}
	\caption{Measured path loss, CI model, FSPL, and 3GPP TR 36.777 UMa-AV model fitting at three frequency bands under different propagation conditions: (a)~LoS, 2.85~GHz; (b)~LoS, 4.6~GHz; (c)~LoS, 7.25~GHz; (d)~NLoS, 2.85~GHz; (e)~NLoS, 4.6~GHz; (f)~NLoS, 7.25~GHz.}
	\label{fig:pathloss}
\end{figure*}

The PL is typically fitted using the close-in (CI) model, which is formulated as

\begin{equation}
PL^{\text{CI}}(f,d)[\text{dB}] = FSPL(f, d_\text{0}) + 10n \log_{10} \left( \frac{d}{d_0} \right) + X_\sigma,
\end{equation}
where $d_\text{0} = 1\text{ m}$ represents the reference distance, $n$ is the PLE, and $X_{\sigma}$ represents the SF term modeled as a zero-mean Gaussian random variable with a standard deviation of $\sigma$, i.e., $X_{\sigma} \sim \mathcal{N}(0, \sigma^2)$.

To quantitatively evaluate the accuracy of the PL model, the root mean square error (RMSE) is employed as a performance metric. It is defined as follows
\begin{equation}
\text{RMSE} = \sqrt{\frac{1}{M} \sum_{i=1}^{M} \left( PL_{i} - \hat{PL}_{i} \right)^2},
\label{eq:rmse}
\end{equation}
where $M$ represents the total number of snapshots, and $\hat{PL}_{i}$ is the estimated value derived from the fitted model.

Fig.~\ref{fig:pathloss} presents the measured PL and the FSPL reference at three frequency bands under different propagation conditions, together with the CI model fitting results and the 3GPP TR~36.777 UMa-AV model predictions~\cite{b36.777}. Overall, for each frequency band, the LoS cases exhibit lower PL than the NLoS cases, which is consistent with the stronger blockage and scattering attenuation under NLoS propagation. In addition, for both LoS and NLoS cases, the overall PL increases with carrier frequency, indicating more severe propagation attenuation at higher frequencies. A noticeable fluctuation of the measured PL can be observed around a Tx-Rx distance of approximately 350~m. This fluctuation is mainly related to the site-specific geometry. Since the UAV remains hovering at an altitude much higher than the surrounding scatterers, the received power variation is mainly affected by the changing local environment around the Rx. As the vehicle moves along the road around the UAV, links with similar Tx-Rx distances may pass through different urban regions, such as densely built and sparsely built areas, leading to significantly different scattering and blockage conditions and therefore large PL variations.

For the CI model fitting, the PLEs under LoS conditions are 2.19, 2.14, and 2.16 at 2.85, 4.6, and 7.25~GHz, respectively, with corresponding SF standard deviations of 4.54, 3.61, and 2.82~dB. Under NLoS conditions, the PLEs increase to 2.57, 2.64, and 2.55, while the corresponding SF standard deviations are 5.43, 4.78, and 2.84~dB. These results show that the fitted PLEs do not exhibit a clear monotonic trend with frequency, whereas the NLoS PLEs are consistently larger than the LoS PLEs. Compared with the measured PL, both the FSPL reference and the 3GPP TR~36.777 UMa-AV model generally underestimate the PL. The RMSE values of the 3GPP model under LoS conditions are 6.26, 4.58, and 4.48~dB at 2.85, 4.6, and 7.25~GHz, respectively. Although the 3GPP TR~36.777 UMa-AV model considers A2G links, its model is still largely based on the conventional terrestrial PL modeling framework. The SF results further show that NLoS propagation generally introduces stronger large-scale power fluctuations than LoS propagation. Consequently, to ensure reliable edge coverage for A2G communications in urban environments, a fading margin exceeding 10 dB should be reserved in the link budget at lower frequency bands.

\subsection{RMS Delay Spread}

\begin{figure}[!t]
	\centering
	\subfloat[]{\includegraphics[width=0.49\columnwidth]{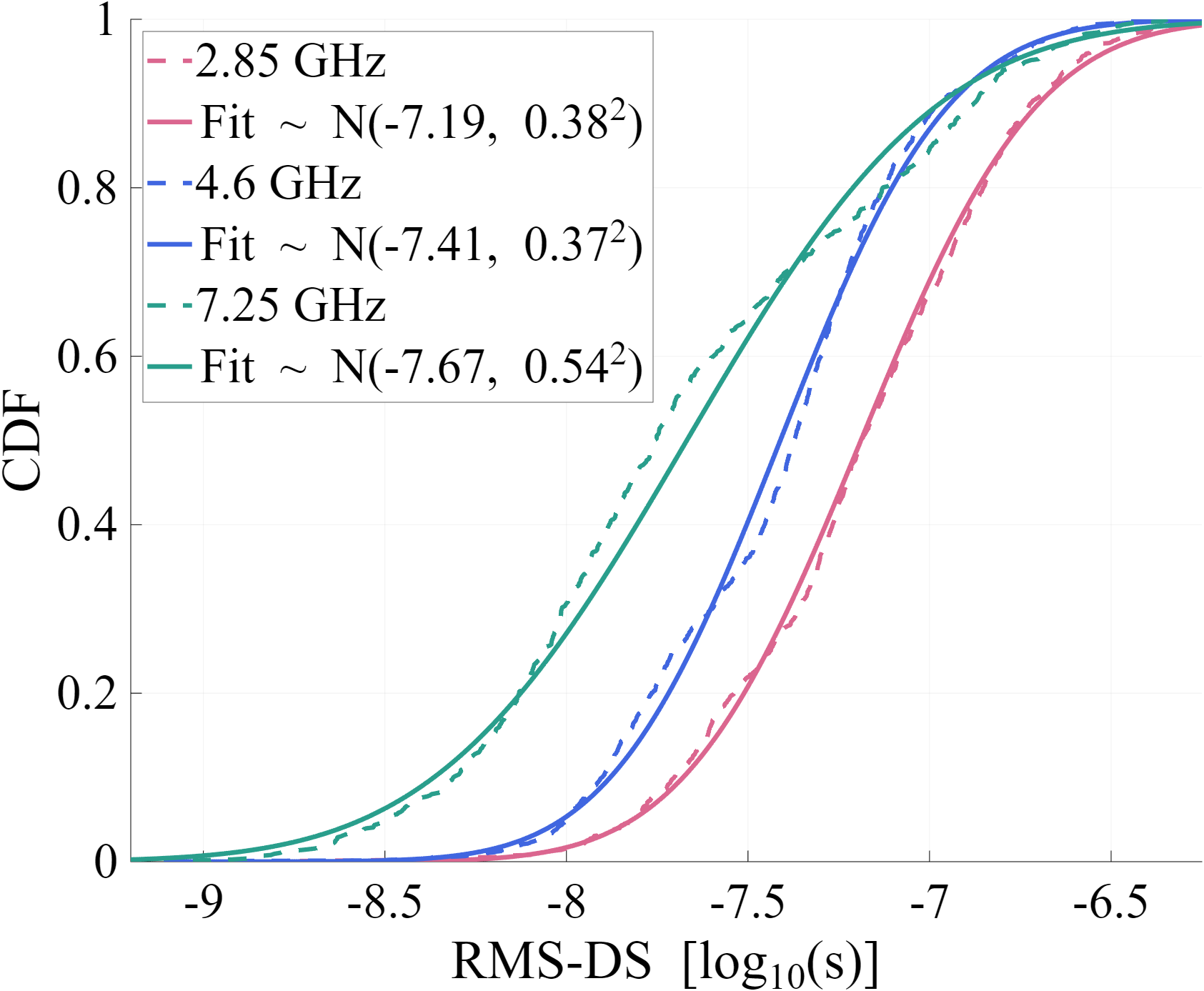}%
		\label{fig:rmsds_los}}
	\hfil
	\subfloat[]{\includegraphics[width=0.5\columnwidth]{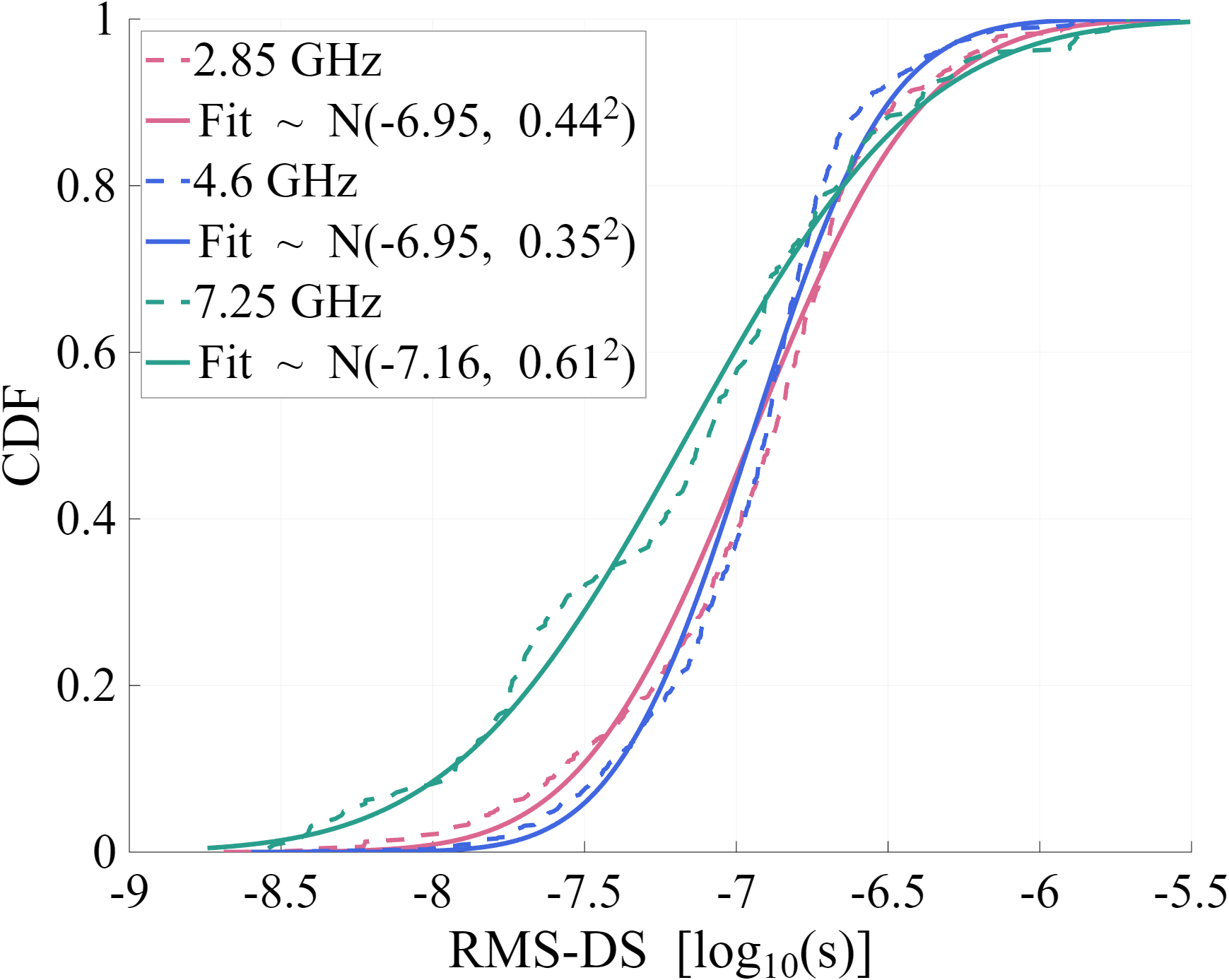}%
		\label{fig:rmsds_nlos}}
	\caption{CDFs and lognormal fitting curves of RMS-DS at 2.85, 4.6, and 7.25~GHz: (a) LoS condition and (b) NLoS condition.}
	\label{fig:rmsds}
\end{figure}

Fig.~\ref{fig:rmsds}(a) and (b) show the measured RMS-DS cumulative distribution functions (CDFs) under LoS and NLoS conditions, respectively, together with the corresponding lognormal fitting curves across the three frequency bands. The fitted parameters are summarized in Table~\ref{tab:large_scale_parameters}. It can be observed that, under LoS conditions, the mean RMS-DS values are 93.11, 54.77, and 46.84~ns at 2.85, 4.6, and 7.25~GHz, respectively, showing a monotonic decrease with increasing carrier frequency. This is because higher-frequency signals generally experience stronger attenuation, making the effective MPCs mainly concentrated around the LoS path. In addition, compared with the LoS cases, the NLoS cases generally exhibit larger mean values and standard deviations of RMS-DS. This difference can be explained by the dominant role of the direct path in LoS propagation, where most of the received power is concentrated around the LoS component and weak delayed MPCs have a limited contribution to the power-weighted delay dispersion. In NLoS propagation, the received power mainly originates from reflected and scattered components, leading to a wider delay domain energy distribution and more pronounced random fluctuations. Compared with the UMa scenario specified in 3GPP TR~38.901~\cite{b38.901}, the measured RMS-DS values in the considered A2G scenario are generally smaller and exhibit a clearer frequency-dependent variation. To characterize the frequency dependence of the RMS-DS statistics, the mean logarithmic RMS-DS values are fitted with respect to the carrier frequency for the LoS and NLoS conditions, respectively. The resulting frequency-dependent models are expressed as

\begin{equation}
	\begin{aligned}
		\mu_{\mathrm{DS,LoS}} &= -6.644 - 1.182\log_{10}(f_c), 
	\end{aligned}
	\label{ds_los}
\end{equation}
\begin{equation}
	\begin{aligned}
		\mu_{\mathrm{DS,NLoS}} &= -6.675 - 0.522\log_{10}(f_c).
	\end{aligned}
	\label{ds_nlos}
\end{equation}

It can be observed that the fitted slope under LoS conditions has a larger magnitude than that under NLoS conditions, indicating that the RMS-DS in LoS links decreases more rapidly with increasing carrier frequency and exhibits a stronger frequency dependence.

From the communication system perspective, a smaller RMS-DS implies weaker frequency selectivity and a larger coherence bandwidth. Therefore, under a given cyclic prefix (CP) overhead constraint, A2G communication systems can support a wider subcarrier spacing (SCS) than terrestrial UMa networks. A wider SCS not only reduces equalization complexity, but also improves the system robustness against Doppler shifts, thereby mitigating inter-carrier interference (ICI) in high-mobility scenarios.

\subsection{Rician $K$-factor}

\begin{table*}[!t] 
	\renewcommand{\arraystretch}{1.4}  
	\renewcommand{\tabularxcolumn}[1]{m{#1}} 
	\caption{Large-Scale Parameters of Multi-Frequency A2G Channels in Urban Scenarios.}
	\begin{center}
		\begin{tabularx}{\textwidth}{>{\hsize=1.3\hsize\centering\arraybackslash}X|%
				>{\hsize=0.7\hsize\centering\arraybackslash}X|%
				>{\centering\arraybackslash}X|>{\centering\arraybackslash}X|>{\centering\arraybackslash}X|>{\centering\arraybackslash}X|>{\centering\arraybackslash}X|>{\centering\arraybackslash}X}
			\hline \hline
			\multicolumn{2}{c|}{Frequency} & \multicolumn{2}{c|}{2.85 GHz} & \multicolumn{2}{c|}{4.6 GHz} & \multicolumn{2}{c}{7.25 GHz} \\
			\hline
			\multicolumn{2}{c|}{Propagation State} & LoS & NLoS & LoS & NLoS & LoS & NLoS \\
			\hline
			PL (dB) & $n$ & 2.19 & 2.57 & 2.14 & 2.64 & 2.16 & 2.55 \\
			\hline
			SF (dB) & $\sigma_{\text{SF}}$ & 4.54 & 5.43 & 3.61 & 4.78 & 2.82 & 2.84 \\
			\hline
			\multirow{2}{*}{\begin{tabular}{@{}c@{}}RMS-DS \\ ($\log_{10}$(1s))\end{tabular}} & $\mu$ & -7.19 & -6.95 & -7.41 & -6.95 & -7.67 & -7.16 \\
			\cline{2-8}
			& $\sigma$ & 0.38 & 0.44 & 0.37 & 0.35 & 0.54 & 0.61 \\
			\hline
			\multirow{2}{*}{KF (dB)} & $\mu$ & 9.16 & N/A & 12.88 & N/A & 12.59 & N/A \\
			\cline{2-8}
			& $\sigma$ & 5.55 & N/A & 5.14 & N/A & 4.21 & N/A \\
			\hline
			\multirow{3}{*}[-1ex]{\begin{tabular}{@{}c@{}}Decorrelation\\Distances\end{tabular}} & SF (m) & 30.68 & 5.09 & 17.58 & 3.35 & 2.11 & 0.79 \\
			\cline{2-8}
			& DS (m) & 12.79 & 1.41 & 12.20 & 1.08 & 1.04 & 0.82 \\
			\cline{2-8}
			& KF (m) & 29.04 & N/A & 11.61 & N/A & 0.89 & N/A \\
			\hline
			\multirow{3}{*}[-1ex]{\begin{tabular}{@{}c@{}}Cross-\\Correlations\end{tabular}} & DS vs SF & 0.33 & 0.04 & 0.34 & 0.09 & 0.28 & 0.13 \\
			\cline{2-8}
			& DS vs KF & -0.50 & N/A & -0.56 & N/A & -0.49 & N/A \\
			\cline{2-8}
			& KF vs SF & -0.61 & N/A & -0.75 & N/A & -0.42 & N/A \\
			\hline \hline
		\end{tabularx}
		\label{tab:large_scale_parameters}
	\end{center}
\end{table*}

\begin{figure}[!t]
	\centering
	\includegraphics[width=0.5\columnwidth]{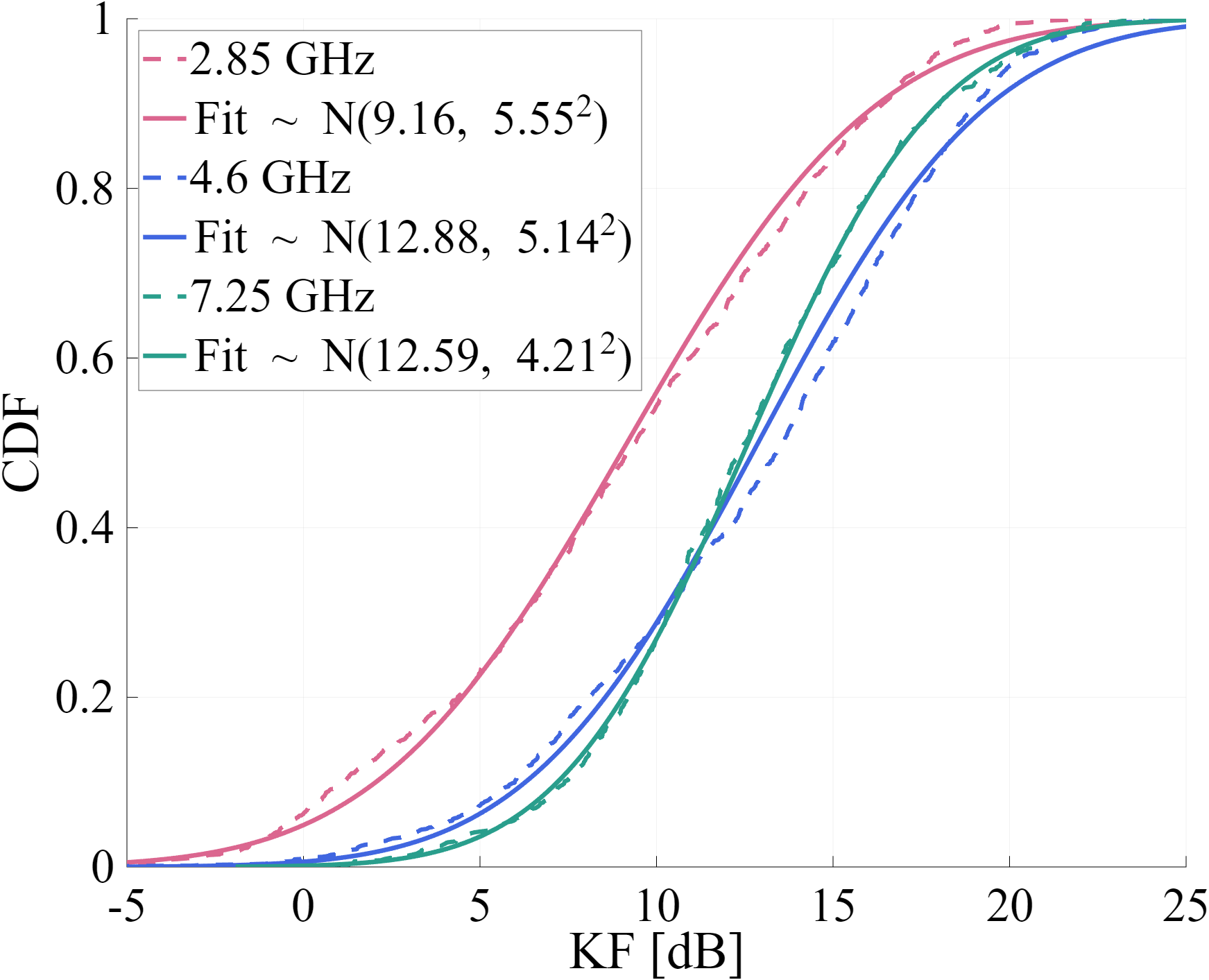}
	\caption{CDFs and normal fitting curves of the Rician $K$-factor at 2.85, 4.6, and 7.25~GHz.}
	\label{fig:KF}
\end{figure}

Fig.~\ref{fig:KF}(a) presents the CDFs of the Rician $K$-factor at the three frequency bands, together with the corresponding normal fitting curves. It can be observed that the Rician $K$-factor values are generally larger than 0~dB, indicating that the LoS path accounts for the majority of the A2G channel energy. Specifically, the LoS component accounts for 89.2\%, 95.1\%, and 94.8\% of the total received channel energy at 2.85, 4.6, and 7.25~GHz, respectively. Similarly, \cite{A2G3p5cam} reported a mean $K$-factor of 15.65~dB at 3.5~GHz in a campus scenario, while \cite{A2G6p5v1} observed that the $K$-factor is mainly distributed between 5.15 and 8.51~dB at 6.5~GHz in a suburban environment. In addition, compared with the UMa scenario specified in 3GPP TR~38.901, where the mean of the $K$-factor remained fix at 9~dB, the $K$-factor in the considered A2G scenario shows an overall increasing trend with carrier frequency. To characterize the frequency dependence of the Rician $K$-factor, the mean $K$-factor values are fitted with respect to the carrier frequency. The resulting frequency-dependent model is given by
\begin{equation}
	\mu_{K}
	=
	8.685*\exp(0.0571*f_c),
	\label{eq:k_factor_freq}
\end{equation}

Moreover, the standard deviations at all three frequency bands exceed the 3.5~dB reference value of the UMa scenario. This suggests that obstacle-induced variations in the measurement environment have a more pronounced impact on the relative strength of the LoS and scattered components.

\subsection{Decorrelation Distance of Large-Scale Parameters}

Traditional research on wireless channels is predominantly founded on the wide-sense stationarity (WSS) assumption, which posits that the statistical properties of the channel remain approximately unchanged within a certain spatio-temporal range. However, in A2G communication scenarios, the relative motion between the Tx and Rx renders the WSS assumption valid only within a limited range. Therefore, accurately characterizing the spatial stationarity of A2G channels is crucial. The spatial stationarity of the channel is evaluated using the autocorrelation coefficients of large-scale parameters, defined as

\begin{equation}
	\rho_X(\Delta d) = 
	\frac{
		E\left[
		\left(X(d)-\mu_X\right)
		\left(X(d+\Delta d)-\mu_X\right)
		\right]
	}{
		\sigma_X^2
	},
	\label{eq:lsp_acf}
\end{equation}
where \(X(d)\) denotes the large-scale parameter value at spatial position \(d\), \(\Delta d\) is the spatial separation distance, \(\mu_X\) and \(\sigma_X\) are the mean and standard deviation of the large-scale parameter sequence, respectively, and \(E\{\cdot\}\) is the expectation operator. In this paper, the decorrelation distance, denoted as $d_c$, is defined as the $\Delta d$ at which the autocorrelation coefficient decays to $1/e$.

Table~\ref{tab:large_scale_parameters} summarizes the decorrelation distances of SF, RMS-DS, and the Rician $K$-factor under different frequency bands and propagation states. Overall, the decorrelation distances of all considered LSPs decrease as the carrier frequency increases. For example, the SF decorrelation distance decreases from 30.68 to 2.11~m under LoS conditions and from 5.09 to 0.79~m under NLoS conditions as the frequency increases from 2.85 to 7.25~GHz. This reduction can be partly attributed to the shorter wavelength at higher frequencies. Under a similar spatial sampling interval, the receiver displacement corresponds to a larger number of wavelengths at higher frequencies, making the channel parameters more sensitive to local spatial variations. In addition, higher-frequency signals have weaker diffraction capability and stronger sensitivity to local blockage and scattering,  which leads to sharper shadow transitions and faster spatial decorrelation of LSPs.

Meanwhile, the NLoS cases generally exhibit shorter decorrelation distances than the LoS cases. This is because LoS propagation is mainly governed by a relatively stable direct path, which helps keep the large-scale channel parameters stable over a longer distance. In contrast, NLoS propagation relies more on reflections, diffraction, and scattering from surrounding buildings and local objects. Therefore, small changes in the Rx position may cause rapid variations in the dominant propagation paths, leading to shorter decorrelation distances in NLoS conditions.

\subsection{Cross-Correlation of Large-Scale Parameters}

In practical wireless channels, LSPs are usually correlated rather than strictly independent. Therefore, the cross-correlations between different LSPs are evaluated to characterize their linear dependence. For two arbitrary LSP vectors, $\mathbf{u}$ and $\mathbf{v}$, each containing (N) samples, the Pearson correlation coefficient is calculated as

\begin{equation}
\rho_{\mathbf{u},\mathbf{v}} = \frac{1}{N-1} \sum_{i=1}^{N} \left( \frac{u_i - \mu_{\mathbf{u}}}{\sigma_{\mathbf{u}}} \right) \left( \frac{v_i - \mu_{\mathbf{v}}}{\sigma_{\mathbf{v}}} \right)
\label{eq:correlation}
\end{equation}
where $u_i$ and $v_i$ denote the LSP samples of the $i$-th snapshot. Table~\ref{tab:large_scale_parameters} summarizes the cross-correlation coefficients of the multi-frequency A2G channels. Under LoS conditions, the KF exhibits a consistent negative correlation with RMS-DS, with correlation coefficients of -0.50, -0.56, and -0.49 at 2.85, 4.6, and 7.25~GHz, respectively. This is physically reasonable because a larger KF indicates that more received power is concentrated in the LoS path, while the contribution of scattered components is reduced, thereby decreasing the delay dispersion and resulting in a smaller DS. A negative correlation is also observed between the KF and SF, with coefficients ranging from -0.42 to -0.75. In addition, RMS-DS and SF show weak positive correlations under both LoS and NLoS conditions, suggesting that delay dispersion is mildly associated with shadow-fading variations.

\section{Conclusion}

In this paper, a multi-frequency wideband A2G channel measurement campaign was conducted in a low-altitude urban scenario at 2.85, 4.6, and 7.25~GHz, each with a bandwidth of 250~MHz. Based on the measured data, a weakly supervised LoS/NLoS propagation state identification method was proposed by integrating geometry-based Fresnel clearance priors, channel features, and spatial consistency constraints. The identified LoS/NLoS state sequences achieve high identification accuracy and exhibit good physical interpretability.

Furthermore, the large-scale and small-scale channel characteristics were extracted and statistically modeled under different frequency bands and propagation conditions. The results show that the path loss generally increases with carrier frequency, whereas the PLEs fitted by the CI model do not exhibit a clear frequency-dependent trend. In addition, the SF follows a normal distribution with standard deviations ranging from 2.82 to 5.43~dB. For link budget design, a fading margin of about 10~dB should be reserved at lower frequency bands to ensure reliable edge coverage. The delay-domain results further show that higher-frequency A2G channels exhibit fewer effective MPCs and weaker delay dispersion, indicating increased channel sparsity. Moreover, the Rician $K$-factor remains generally larger than 0~dB and can be well characterized by a normal distribution.

Finally, the cross-correlation and decorrelation-distance characteristics of large-scale parameters were evaluated across different frequency bands and propagation conditions. The cross-correlation analysis shows that the Rician (K)-factor is negatively correlated with both SF and RMS-DS under LoS conditions, indicating that a stronger dominant LoS component is usually accompanied by weaker shadowing fluctuation and smaller delay dispersion. In terms of spatial stationarity, the SF decorrelation distances under LoS conditions are 30.68, 17.58, and 2.11~m at 2.85, 4.6, and 7.25~GHz, respectively, showing that the spatial correlation of large-scale parameters becomes weaker at higher frequencies. The NLoS cases generally exhibit shorter decorrelation distances than the LoS cases, because the received signal relies more strongly on local reflections, diffraction, and scattering. These results indicate that urban low-altitude A2G channels exhibit pronounced spatial non-stationarity, which should be carefully considered in future 6G low-altitude channel modeling and system design.


\end{document}